\documentclass[
journal=jcisd8, 
manuscript=article,
layout=twocolumn   
]{achemso}

\usepackage[T1]{fontenc} 

\usepackage{graphicx}
\graphicspath{ {./Figures4Submission/} }

\usepackage{pbox}
\usepackage{tabularx}

\usepackage{mathtools} 
\usepackage{amssymb} 
\usepackage[flushleft]{threeparttable}

\usepackage{dcolumn}
\usepackage{bm}
\usepackage[normalem]{ulem}
\usepackage{xcolor}
\usepackage{hyperref}

\definecolor{OliveGreen}{rgb}{0,0.6,0}
\definecolor{auburn}{rgb}{0.43, 0.21, 0.1}
\definecolor{BlueViolet}{rgb}{0.54, 0.17, 0.89}
\definecolor{HokieOrange}{RGB}{232, 119, 34}
\definecolor{HokieMaroon}{RGB}{134, 31, 65}
\definecolor{Black}{RGB}{0, 0, 0}

\usepackage{appendix}
\usepackage{chngcntr}
\usepackage{etoolbox}
\usepackage{lipsum}
\usepackage{makecell}
\usepackage{natbib}
\usepackage{subcaption}

\author{Jared McDonald}
\affiliation{Materials Science \& Engineering Department, Virginia Tech, Blacksburg, VA 24061, USA}
\email{jmcdonald@vt.edu}

\author{Michael R. von Spakovsky}
\affiliation{Mechanical Engineering Department, Virginia Tech, Blacksburg, VA 24061, USA}
\email{vonspako@vt.edu}

\author{William T. Reynolds Jr.}
\affiliation{Materials Science \& Engineering Department, Virginia Tech, Blacksburg, VA 24061, USA}
\email{reynolds@vt.edu}

\title{Predicting Ion Sequestration in Charged Polymers with the Steepest-Entropy-Ascent Quantum Thermodynamic Framework}

\begin{document}




\begin{abstract}
The steepest-entropy-ascent quantum thermodynamic framework is used to investigate the effectiveness of multi-chain polyethyleneimine-methylenephosphonic acid in sequestering rare-earth ions (Eu$^{+3}$) from aqueous solutions. The framework applies a thermodynamic equation of motion to a discrete energy eigenstructure to model the binding kinetics of europium ions to reactive sites of the polymer chains.
The energy eigenstructure is generated using a non-Markovian Monte Carlo model that estimates energy level degeneracies. The equation of motion is used to determine the occupation probability of each energy level, describing the unique path through thermodynamic state space by which the polymer system sequesters rare-earth ions from solution. A second Monte Carlo simulation is conducted to relate the kinetic path in state space to physical descriptors associated with the polymer, including the radius of gyration, tortuosity, and Eu-neighbor distribution functions. These descriptors are used to visualize the evolution of the polymer during the sequestration process. The fraction of sequestered Eu$^{+3}$ ions depends upon the total energy of the system, with lower energy resulting in higher sequestration. The kinetics of the overall sequestration are dependent on the steepest-entropy-ascent principle used by the equation of motion to generate a unique kinetic path from an initial non-equilibrium state.
\end{abstract}

\maketitle

\section{Introduction}

Rare earth elements play a crucial role in many electrical and electronic devices, and their importance has only grown in recent times. However, concerns have been raised about the future availability of these metals because of their limited recycle-ability, geopolitical constraints on the supply chain, and current consumption rates. As a result, there is a growing interest in exploring alternative sources such as seawater and effluent streams from which these elements can be extracted~\cite{Diallo2015, Archer2020, Sener2021, Barakat2010}. One method for recovering rare earth elements from solutions is through polymer-ion complexation, also known as sequestration or chelation. This approach utilizes charged components of a polymer chain or matrix to trap rare earth ion complexes present in a polar solution. Experimental and computational studies have explored various classes of polymers and resins with diverse structures for this approach~\cite{Abderrahim2006, Archer2020,  Archer2021, Archer2022Resins, Hu2017, Adel2010, Fllorek2015}. Research in this area has also extended to ion transport in liquid electrolytes for energy storage applications~\cite{Wang2016, Ketkar2019}.

A polymer that has received particular attention for sequestration is polyethyleneimine (PEI). Interest in PEI stems from its commercial availability and its ability to be tailored to specific chemical applications. Its versatility comes from the ability to add functional groups selectively to interact with specific metallic species~\cite{Archer2020, Zhao2020, Abderrahim2006, Zhao2020}. Polyethyleneimine applications include sequestration of rare earth and radioactive elements, radionuclide transport for cancer treatment, and gene editing~\cite{Sun2011, Andreoli2014, Popwell2014, Zhu2015}. Among the various functional groups that can be used for sequestration, methylenephosphonic acid is a popular choice. Functionalized methylenephosphonic acid is effective for collecting metals, and when combined with PEI, it surpasses other groups for sequestering rare earth elements~\cite{Archer2022Resins, Popwell2014, Zhu2015, Abderrahim2006,Archer2020}. The phosphonated form of PEI is known as poly(ethylenimine methylene phosphonate), or PEI-MP. 

Although PEI-MP systems have been studied experimentally and computationally, the binding thermodynamics are not understood well. Experimental investigations of sequestration systems have employed isothermal titration calorimetry to measure heat transfer during polymer ion-binding events and assess enthalpic changes during the sequestration process. This technique has been used to correlate thermodynamic properties to the sequestration performance of a variety polymer systems~\cite{Archer2020, Termuhlen2017, Price2018, Archer2022Resins}. The underlying thermodynamic driving force for sequestration has been studied computationally~\cite{Schmidt2002,Welborn2023}. Isothermal titration calorimetry indicates a reduction of entropy drives ion sequestration\cite{Archer2020}. This finding should have implications on computational modeling of sequestration efficiency and system dynamics, but computational models like molecular dynamics do not explicitly consider entropy.  Moreover, there is a lack of information about the spatial arrangement of ions within PEI-MP during complexation. In most of the literature, the structure of PEI-MP takes the form of a branched resin, but a recent report has considered a linearized form of PEI-MP~\cite{Archer2020}. Application of this new form is intended to increase the mobility of the chains and improve the system's chelation efficiency~\cite{Archer2020}, but there is no analysis of this PEI-MP structure beyond the experimentally predicted number of pair interactions between the ions and functional groups.

These circumstances suggest that a strategy capable of treating entropic effects explicitly and including information about polymer chain conformations could be useful for understanding sequestration. The steepest-entropy-ascent quantum thermodynamics (SEAQT) framework is a promising approach that can address these effects. In this contribution, SEAQT is applied to an energy eigenstructure (a set of discrete internal energy levels and their corresponding degeneracies) to predict the interplay between the thermodynamics, kinetics, and structure of a sequestration system. To generate the energy eigenstructure for a PEI-MP system, we use the Wang-Landau algorithm, a non-Markovian Monte Carlo model that estimates the density of states of a discrete system. We then apply an equation of motion derived from the principle of maximum entropy ascent to predict the kinetics of sequestration. The state of the system at each instant of time is characterized by a set of occupation probabilities for all the energy levels of the system, and the time-evolution of these probabilities provides a deterministic path through the state space of the sequestration process. We also propose a set of descriptors that link this path in state space to physical conformations of PEI-MP and the positions of a rare-earth ion, Eu$^+3$, during sequestration.

The following sections present a thermodynamic model for the discrete energy eigenstructure of PEI-MP in a polar solvent containing Eu$^{+3}$ (Section~\nameref{SectionIIA}) and a description of the Wang-Landau algorithm (Section~\nameref{SectionIIB}). The SEAQT equation of motion is then introduced (Sections~\nameref{SectionIIC} and \nameref{SectionIID}), along with a method for linking the thermodynamic state space to polymer conformations and sequestration (Section~\nameref{SectionIIE}). Predicted results for the sequestration process along four representative non-equilibrium thermodynamic paths are then presented (Section~\nameref{SectionIII}) followed by a discussion of these results  (Section~\nameref{Discussion}). Finally, some conclusions are given (Section~\nameref{Conclusions}).

\section{Method}
\subsection{Energetic Model}\label{SectionIIA}

We start by developing a molecular model for the energy of a given number of PEI-MP chains in a polar solvent (namely, water) that contains Eu$^+3$ ions. The energy of a particular conformation of molecular chains and the distribution of the ions is taken to be the sum of the pair-wise interactions among all the component molecules.  The molecular arrangement is described with a 3-dimensional cubic lattice using the Bond Fluctuation Model~\cite{Carmesin1988, Deutsch1991}. This model regulates the allowable distance between constituents by tracking the occupation of a given lattice site using a cubic unit cell. Specifically, if a monomer occupies a site at the center of a cube, then other monomers cannot occupy any of the eight nearest-neighbor sites at the corners of the cube. This approach makes it possible to simulate multiple bond lengths between sequential monomers of a polymer chain while not violating the excluded volume principal~\cite{Carmesin1988, Deutsch1991}.

The molecular constituents include the linear PEI chain's monomers, the functionalized phosphonic acid that is attached to the PEI backbone, the Eu ions, and the components of the polar solvent (i.e., the H$_2$O), which we simplify as an oxygen atom with two hydrogen atoms attached at a $90^\circ$ angle on a cubic lattice.

\begin{figure}[ht!]
	\begin{center}
		\includegraphics[width=0.47\textwidth]{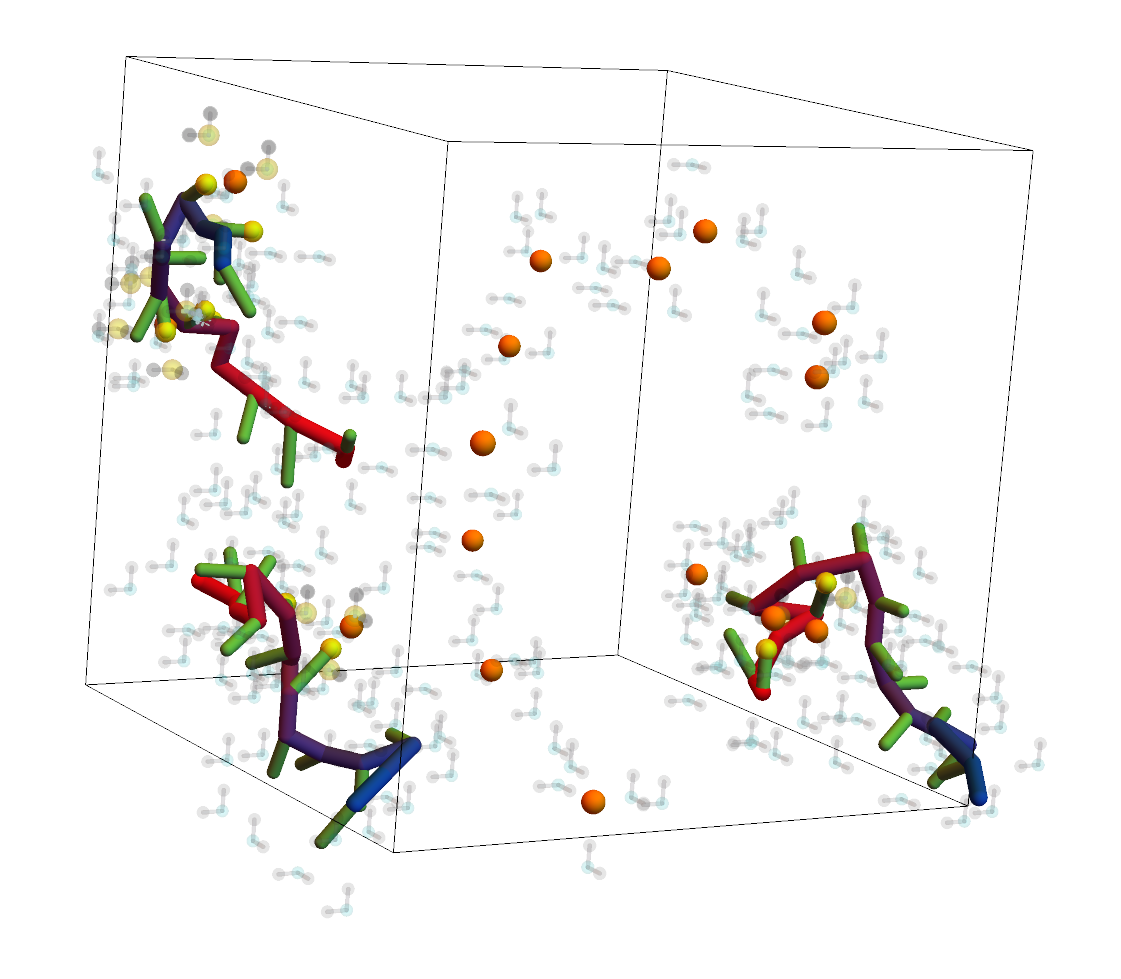}
		\caption{Example microstructure of three polyethylenimine methylene phosphonate (PEI-MP) chains in a solvent of water containing thirty Eu$^{+3}$. The PEI chains are shown as tubes with a red-blue color gradient to help distinguish their orientation. The water molecules are represented by faint grey H atoms attached to a cyan O atom; only a small portion of the 900 water molecules in the simulation domain are shown to avoid obscuring the PEI-MP chains. Free Eu$^{+3}$ ions are shown as orange spheres and sequestered Eu ions that are captured by the functional MP groups and the oxygen atoms in H$_2$O are represented by yellow spheres.}
		\label{ExampleSystemMicrostructure}
	\end{center}
\end{figure}

The simulation domain is a cube with periodic boundary conditions on all sides. The length of each edge is $30 \, a$ where $a$ is the lattice parameter of the Bond Fluctuation Model. An example of one arrangement of the simulation domain is shown in Figure~\ref{ExampleSystemMicrostructure}. It includes 3 PEI chains (each 15 monomers long), 30 Eu$^+3$ ions, and 900 H$_2$O molecules (most of which are not shown in the figure to avoid obscuring the PEI chains). The methyl phosphonate functional groups are represented as the short segments protruding from the PEI backbone. Each chain has 86\% of its monomers functionalized with attached methyl phosphonate groups. To avoid nonphysical bond cutting, we limit the length of the bonds between monomers of the polymer chain to the values $\sqrt{4}a$, $\sqrt{5}a$, $\sqrt{6}a$, $\sqrt{8}a$, $\sqrt{9}a$, or $\sqrt{10}a$. The atoms of the water molecules are spaced one lattice parameter apart, whereas the spacing of all other bonded constituents is $2a$ apart.

The energy of a specific conformation of the polymer chains, water molecules, and Eu ions can be determined by summing the pairwise interaction potentials between each molecule and all the other molecules in the simulation domain. The pairwise potentials depend upon the identity of the two species of the interacting pair. The discrete energy, $E_j$, of a particular configuration is given by the expression

\begin{equation}
E_j = \frac{1}{2}\sum_{n=1}^{N_{\text{tot}}} \left( \sum_{\substack{m=1 \\ m\neq n}}^{N_{\text{tot}}} V_{n,m}^{\phi} \right)
\label{eqn:Ej}
\end{equation}
where $m$ and $n$ correspond to the components of an interacting pair. The summations are over the $N_{tot}$ atoms/molecules in the system. The pair potential of the interacting components is represented by $V^{\phi}_{n,m}$. The $\phi$ superscript switches the intermolecular potential according to the identity of the bonding pairs: an intra-chain potential for bonding along the polymer chain, a Lennard-Jones potential for repulsive van der Waals interactions, and an electrostatic potential for Coulombic interactions between ions and polar parts of molecules. All the possible interactions and the constants employed in the potentials are listed in Table~\ref{PairPotentials}. 

\begin{table*}[htbp]
\caption{Pair potential functions used in the bond fluctuation model for the PEI-MP system.}
\label{PairPotentials} 
 \begin{threeparttable}
    \centering
\footnotesize
\begin{tabularx}{0.99\textwidth} { 
  | >{\raggedleft\arraybackslash}c 
  | >{\raggedright\arraybackslash}c 
  | >{\raggedright\arraybackslash}X 
  | >{\raggedright\arraybackslash}c| }
  \hline
 & \multicolumn{1}{|c|}{$n,m$} & \multicolumn{1}{|c|}{pair} &  \\
 & \multicolumn{1}{|c|}{molecule} & \multicolumn{1}{|c|}{potential} &  \\ 
\multicolumn{1}{|c|}{ $V^{\phi}_{n,m}$ } & \multicolumn{1}{|c|}{pair types} & \multicolumn{1}{|c|}{function} & \multicolumn{1}{|c|}{parameters} \\[2mm] \hline
$V_{\text{LJ}}$ & \makecell{van der Waals \\ PEI : PO : Eu : H : O} & \makecell{$\; 4 \varepsilon _{\alpha\beta}\left(\left(\frac{\sigma }{r_{\text{}}}\right)^{12}-\left(\frac{\sigma }{r_{\text{}}}\right)^6\right)$ }& \makecell{$\sigma=3.3$ \AA,\\$\varepsilon _{\text{PEI-PEI}}/k_b =392K$\\$\varepsilon _{\text{PO-PO}}/k_b=265K$\\$\varepsilon _{\text{Eu-Eu}}/k_b=58K$\\$\varepsilon _{\text{H-H}}/k_b=405K$\\$\varepsilon _{\text{O-O}}/k_b=405K$} \tnote{a}
\tnote{b} \tnote{c} \rule[-.3\baselineskip]{0mm}{9mm}

\\$V_{\text{Bond}}$ & intra-chain monomer  & \makecell{$\;\; K_b\frac{1}{2}(r_{eq}-r)^2$\\$\;$} \rule[-.3\baselineskip]{0mm}{9mm}& $K_b=1000, r_{eq}=3.0$ \AA \tnote{d} \rule[-.3\baselineskip]{0mm}{5mm} 

\\$V_{\text{Elec}}$ & \makecell{Coulomb \\ PEI : PO : Eu : H : O}&
\makecell{$\frac{{q_1}{q_2}}{4 \pi \epsilon _{0} r} ,\; r\leq a\sqrt{13}$ \\
$(\frac{{q_1}{q_2}}{4 \pi \epsilon _{0} 4}+\frac{{q_1}{q_2}}{4 \pi \epsilon _{0} 15})/2 ,\;  r> a\sqrt{13}$}
\tnote{e}
 & \makecell{$q_{\text{PEI}} = .435,q_{\text{PO}} =-.834 ,$\\$q_{\text{Eu}} = 3,q_{\text{O}} = -.834,q_{\text{H}} =.435 $}  \tnote{f}
\tnote{g} \rule[-.3\baselineskip]{0mm}{9mm}   \\
\hline
\end{tabularx}

\begin{tablenotes}
  \item[a] Values are from reference \cite{Wilhelm1971,Poling2001,Li2015} and $k_b$ is Boltzmann's constant.
  \item[b] Mixed particle interactions are approximated using the Lorentz and Berthelot equation $\varepsilon _{\alpha\beta} =\sqrt{\varepsilon _{\alpha\beta}\;\varepsilon _{\alpha\beta}}$.
  \item[c] Interaction parameters for H$_2$O molecule are halved between H$_2$ and O and halved between the two H atoms.
  \item[d] Reference values from \cite{Rossi2011} for $r_{\text{eq}}$ are increased to reduce the computational impact of the functions exponential energy change.
  \item[e] Long-range electrostatic interactions are represented as an averaged value for computational efficiency.
  \item[f] Values are from references \cite{Jorgensen1983,Laury2015}.
  \item[g] Electrostatic charges resulting from similar atomic sources are considered the same.
\end{tablenotes}
\end{threeparttable}

\end{table*}

For computational efficiency, one value of the Lennard-Jones parameter, $\sigma=3.3$ {\AA } and one bond length $r_{\text{eq}}= 3.0$ {\AA } was used for all the constituents. Although this simplification can be expected to affect the accuracy of the energy eigenstructure, qualitative relationships among parameters, like $r_{\text{eq}}< \sigma$, are preserved. The simplification also ensures all Lennard-Jones parameters are negative over the full interaction range. Because $\sigma$ sets the length scale of the majority of interactions, the lattice parameter ($a=1.85$\AA) is inferred from the relationship, $2a=2^{1/6}\sigma=3.7$ \AA.

\subsection{Wang-Landau Algorithm}\label{SectionIIB}

The Wang-Landau algorithm is a powerful non-Markovian Monte Carlo method for estimating the density of states, or degeneracies, $g(E_j)$, associated with the discrete energy levels of a system~\cite{Li2014, Vogel2013, WangLandau2001a, WangLandau2001b}. 
It has been used to calculate stable equilibrium thermodynamic properties~\cite{Wust2012, Farris2019}, and it has been used to build the energy eigenstructures for metallic, ceramic, semiconducting, and polymeric materials~\cite{McDonald2022,Younis2022,McDonald2023polymer}. The Wang-Landau algorithm estimates the density of states by randomly walking through the discrete energy levels in a way that ensures all the levels are sampled uniformly~\cite{LandauTsaiExler2004}.  Under this condition, the sampling probability of visiting a particular energy level, $p_j^s$, is proportional to the inverse of the level's degeneracy: $p_j^s \propto 1/g_j$. This relationship can be used to estimate all the $g_j$ values from a Monte-Carlo simulation provided the whole spectrum of energy levels is uniformly sampled. The Wang-Landau algorithm, and parallelized variants called Replica Exchange Wang-Landau~\cite{Li2014, Vogel2013, WangLandau2001a, WangLandau2001b, LandauTsaiExler2004, Wust2012, Farris2019, Vogel2013, Vogel2014, Vogel2018} estimate the $g(E_j)$ for a system within a quantifiable degree of uncertainty.

\begin{figure}[ht!]
	\begin{center}
			\includegraphics[width=0.47\textwidth]{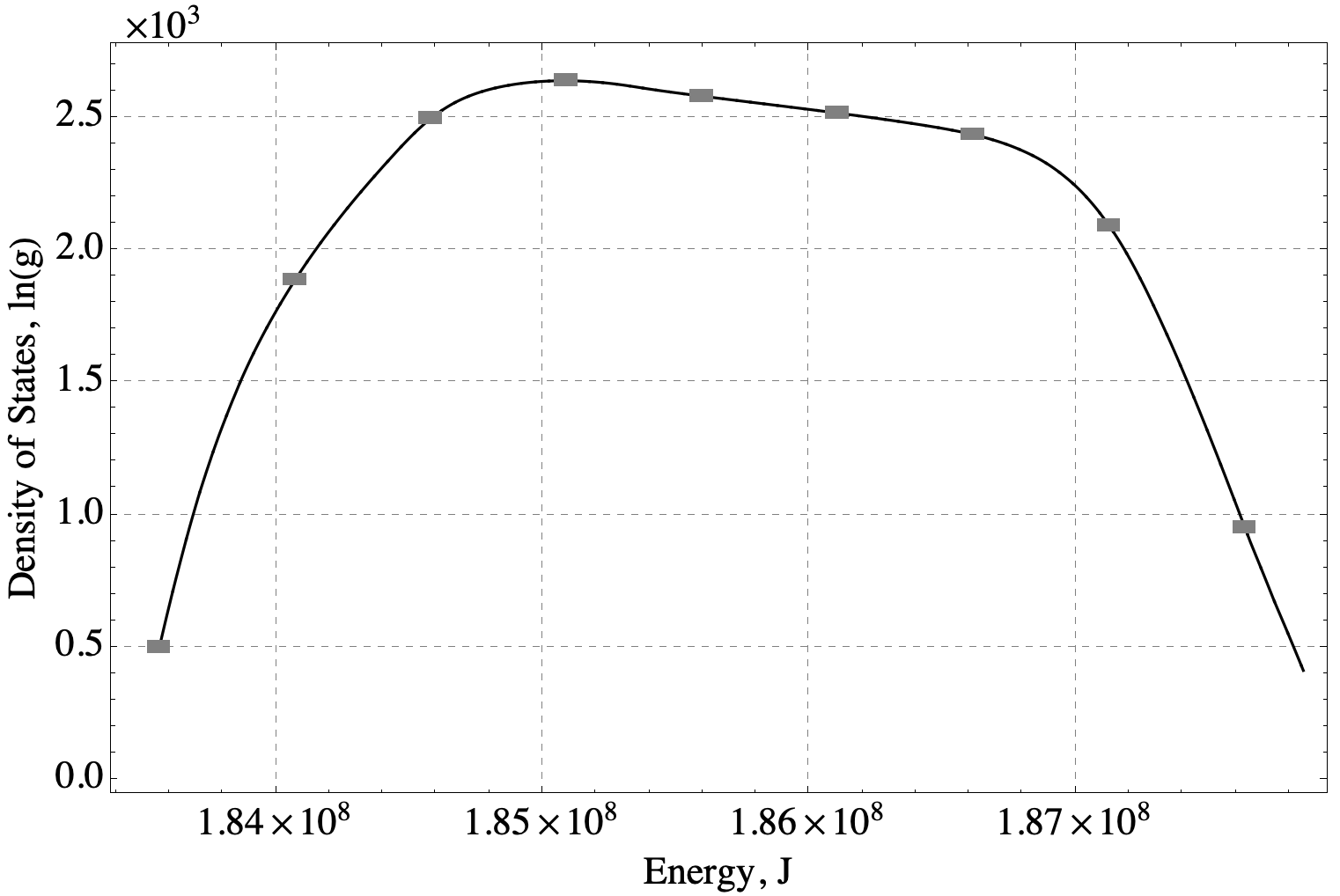}
		\caption{The energy eigenstructure estimated with the Replica Exchange Wang-Landau algorithm for the PEI-MP sequestration system illustrated in Figure~\ref{ExampleSystemMicrostructure}. The natural logarithm of the degeneracy, $\ln(g)$, of the $j^{th}$ energy eigenlevel versus the energy, $E_j$, of each level. The degeneracy is estimated for 3 PEI chains, 30 Eu$^+3$ ions, and 900 H$_2$O molecules in a simulation domain $30a$ on each side. The $E_j$ are calculated from summed pairwise interaction energies.}
		\label{DOS}
	\end{center}
\end{figure}

For optimal results, the Wang-Landau method should be applied in a way that is tailored to the physical system being studied. The number of energy levels in the system should be matched with the available computational resources for calculating the degeneracies. For instance, the degeneracies for a system with $10^5$ discrete energy levels can be calculated in less than 72 hours using a desktop workstation such as an iMac with 32 GB of RAM and 16 Ryzen 9
5800X processors. To change the number of energy levels, one approach is to scale the energy model used to calculate the energy of a particular conformation (Equation~\ref{eqn:Ej}) in such a way that the precision of the energy level values is adjusted. This can be done by multiplying the parameters used in the interatomic potentials by an appropriate multiple of 10 and truncating the remaining decimal. Alternatively, if the possible configurations or conformations of the system produce many closely-spaced energy levels, they can be binned to generate a coarse-grained, pseudo-eigenstructure for the system. Both scaling the energy precision and binning the energy levels can increase the speed with which the degeneracies are generated, especially for energy levels at the low-end of the energy spectrum.

Figure~\ref{DOS} shows the density of states generated with the Replica Exchange Wang-Landau algorithm from Equation~\ref{eqn:Ej} for the system under consideration, which consists of 3 PEI chains, each 15 monomers long, 30 Eu$^+3$ ions, and 900 H$_2$O molecules. The energy eigenstructure has 5800 energy levels, and the degeneracy of each level depends on its energy. The overall shape of the curve for the logarithm of the energy level degeneracy, $\ln(g)$, versus $E_j$, depends on the nature of the physical system. Low degeneracies at low energies are typical~\cite{McDonald2022,McDonald2023polymer,Younis2022}, but the low degeneracies for the highest energy levels shown in Figure~\ref{DOS} are less common. These lower degeneracies may be related to the longer-range interactions that arise from the electrostatic interactions in the  energy model for the sequestration system. 

The error bars in Figure~\ref{DOS} represent relative error scaled from a statistical analysis of a similar polymer system~\cite{McDonald2023polymer}. The density of states of a 58--monomer chain was estimated with six independent trials using the same Wang-Landau parameters as those used for the sequestration system, and error bars representing $\pm$ one standard deviation of the degeneracy were determined from the six simulations~\cite{McDonald2023polymer}.  Because the degeneracy of the single-chain system and the sequestration system are different, the error bars in Figure~\ref{DOS} were scaled to represent the same relative error expected in the sequestration system.

Finally, another statistical approach (e.g., multicanonical or replica-exchange Monte Carlo) could have been used to determine the system energy levels and associated degeneracies, but in the end the Replica Exchange Wang-Landau algorithm was selected for its balance of accuracy and computational efficiency \cite{Farris2021,Hayashi2019}. In fact, Farris and Landau \cite{Farris2021} have shown that this algorithm compares favorably even at the lowest energy levels (i.e., typically the most difficult to resolve) with PERM \cite{Thachuk2007}, which is one of the best multicanonical methods, and with REMC, which is a replica-exchange Monte Carlo method \cite{Swendsen1986,Geyer1991,Hukushima1996}. In addition, the detailed balance condition required by these other methods for convergence to the density of states is not required for convergence by the Wang-Landau algorithm  \cite{Wust2012,Zhou2008}. However, it is required for the trial moves used in the algorithm so that systematic errors are avoided. This is accomplished using a trial-and-error procedure that randomly chooses trial moves independently of the current conformation. It does this with constant probability \cite{Wust2012}. In addition, use of the Wang-Landau algorithm is limited here to predicting the system's energy spectrum and associated degeneracies. Both of these are utilized by the SEAQT equation of motion, which is described in the next section, to predict the kinetics of each non-equilibrium path based on the principle of steepest entropy ascent. The non-equilibrium states predicted by this equation do not require a detailed balance, and it is even questionable to suppose such a balance holds at non-equilibrium. Of course, because the Wang-Landau and Replica Exchange Wang-Landau algorithms are approximate as indeed all similar statistical methods are, there is uncertainty in the energy levels and degeneracies that they predict, particularly at the low energy levels. Nonetheless, granting these unavoidable uncertainties, they do not significantly affect the results predicted by the SEAQT equation of motion because the lowest energy levels contribute little to the non-equilibrium thermodynamic states and paths predicted here.

\subsection{SEAQT Equation of Motion}\label{SectionIIC}

Once the energy eigenstructure of the system is established, how the system progresses from any arbitrary initial state to a stable one can be established deterministically using the SEAQT framework. The kinetic path is predicted with the principle of steepest-entropy-ascent, which is analogous to the principle of maximum entropy production at each instant of time along the path. This framework is unique in the context of kinetic theories in that it does not require an $a \; priori$ specification of the rate-limiting processes or the underlying kinetic mechanisms associated with the kinetic phenomena. Instead, it employs an equation of motion, which is based on a variational principle in thermodynamic state space, to predict a non-equilibrium thermodynamic path from some initial (non-equilibrium) state to one at stable equilibrium~\cite{Beretta1984, Li2016b, Li2018, Li2018steepest, yamada2019spin,yamada2020kineticpartII}. This path is uniquely determined, and because it is in state space, it is not susceptible to getting ``trapped'' in local metastable conformations.  The path is guaranteed to reach the stable equilibrium state corresponding to maximum entropy (or equivalently, minimum energy) and conserve those quantities that must be conserved (e.g., the occupation probabilities, the energy, etc.). In its most general form, the SEAQT equation of motion for a simple quantum system is expressed in operator form as:

\begin{equation}
\frac {d \hat{\rho}} {dt}=\frac {1} {i \hbar}[\hat{\rho},\hat{H}]+\frac {1} {\tau(\hat{\rho})} {\hat D(\hat{\rho})} \label{EOM1}
\end{equation}
where $t$ is the time, $i$ the imaginary unit, $\hbar$ Planck's modified constant, and $\hat{\rho}$ the density operator or so-called ``state operator'', which for the classical system
modeled here is interpreted as a probability distribution. The quantity, $[\hat{\rho},\hat{H}]$, is the commutator of $\hat{\rho}$ and $\hat{H}$, and $\hat{H}$ is the Hamiltonian operator. The first term to the right of the equal sign is the von Neumann term of the time-dependent part of the Schr\"odinger equation of motion for quantum systems. It predicts the purely reversible energy evolution of zero-entropy (i.e., pure) states. The second term on the right side of Equation~\ref{EOM1} is the dissipation operator $\hat D(\hat{\rho})$ multiplied by the inverse of the relaxation parameter, $\tau$.  This additional term accounts for the evolution of the state of the system when there are nonzero entropy states and the process is irreversible~\cite{Beretta2006, Beretta2009, Li2016d}.

Since for classical systems there are no quantum correlations, the first term on the right in Equation~\ref{EOM1} disappears because $\hat{\rho}$, which is diagonal in the energy eigenvalue basis, commutes with $\hat{H}$ \cite{Li2016a,Li2016b,Li2018,Beretta2006,Beretta2009,Li2016d}. The remaining term, the dissipation term, is derived from a variational principle along a gradient descent in Hilbert space in the direction of steepest entropy ascent. For a classical system in which the only generators of the motion are the Hamiltonian and identity operators, the equation of motion takes the non-operator form: \cite{Beretta1984, Beretta1985, Li2016d}

\begin{equation}
\frac{dp_j}{dt}=\frac {1} {\tau}\frac{\left|
\begin{array}{ccc}
 -p_j \ln \frac{p_j}{g_j} & p_j & {E}_j\, p_j \\
 \langle S \rangle & 1 & \langle E \rangle \\
 \langle E\,S \rangle & \langle E \rangle & \langle E^2 \rangle \\
\end{array}
\right|}{\left|
\begin{array}{cc}
 1 & \langle E \rangle \\
  \langle E \rangle & \langle E^2 \rangle \\
\end{array}
\right|} 
\label{EOM2}
\end{equation}
In this equation, the change in the time step $dt$ of the occupation probability, $p_j$, of the $j^{th}$ energy level is given by the ratio of two determinants that depend upon the occupation probability and degeneracy of the level as well as the expectation values of the thermodynamic properties indicated by the quantities in angle brackets, $\langle \, \cdot \, \rangle\;$.  The entropy of the $j^{th}$ energy level follows from the von Neumann definition of the entropy~\cite{Gyftopoulos1997} which takes the form, $S_j = -\ln \frac{p_j}{g_j}$. The expectation values in Equation~\ref{EOM2} are defined as
\begin {align}
\langle E\, \rangle =&\; \underset {j} {{\sum}}\phantom{l} p_j(E_{j}) \label{ExpectationE} \\
\langle E^2\, \rangle =&\; \underset {j} {{\sum}}\phantom{l} p_j(E_{j}^{2})  \\
\langle S\, \rangle =&\; \underset {j} {{\sum}}\phantom{l} p_j(S_{j}) = \underset {j} {{\sum}}\phantom{l} - p_j( \ln \frac{p_j}{g_j}) \label{ExpectationS} \\
\langle E\,S\, \rangle =&\; \underset {j} {{\sum}}\phantom{l} - p_j(E_j \ln \frac{p_j}{g_j}) \label{ExpectationValues}
\end {align}

The form of the equation of motion given by Equation~\ref{EOM2} is applicable to a thermodynamically isolated system, i.e., one in which the total energy remains constant since no energy interactions across the system boundary occur. However, using the hypo-equilibrium concept developed by Li and von Spakovsky \cite{Li2016a} to represent each non-equilibrium state of such a system, its overall Hilbert space of all states can be subdivided into subspaces, which in the case of a non-reacting system, coincide with a subdivision of the isolated system into the same number of non-isolated subsystems  exchanging mass and energy as there are subspaces. For example, assuming that each non-equilibrium state can be described by a $2^\text{nd}$--order hypo-equilibrium state, the isolated system becomes the composite of two non-isolated subsystems $A$ and $B$ that evolve by exchanging energy, or energy and mass, with each other. For the former case, the equation of motion for subsystem $A$ (subspace $A$) is written as:

\begin{equation}
\frac{dp_j^A}{dt}=\frac {1} {\tau}\frac{\left|
\begin{array}{cccc}
 -p_j^A \text{ln}\frac{p_j^A}{g_j^A} & p_j^A &0 & {E}_j^A p_j^A \\
 \langle S^A \rangle & 1 & 0 &\langle E^A \rangle \\
 \langle S^B \rangle & 0 & 1 &\langle E^B \rangle \\
 \langle E\,S \rangle & \langle E^A \rangle & \langle E^B \rangle & \langle E^2 \rangle \\
\end{array}
\right|}{\left|
\begin{array}{ccc}
 1 & 0&
\langle E^A \rangle \\
  0&1&\langle E^B \rangle \\
   \langle E^A \rangle &\langle E^B \rangle &\langle E^2 \rangle\\
\end{array}
\right|}
\label{ABEqM}
\end{equation}

A similar equation of motion is written for subsystem $B$ (subspace $B$). Expressing the determinant of the numerator in Equation~\ref{ABEqM} as an expansion of its first row with co-factors, $C_1$, $C^{A}_{2}$, $C_3$, Equation~\ref{ABEqM} reduces to

\begin{align}
\frac {dp_j^A}{dt^*} & = p_j^A \left(-\text{ln}\frac{p_j^A}{g_j^A}-\frac{C_2^A}{C_1}-{E}_j^A\frac{C_3}{C_1}\right) \nonumber \\
&   = p_j^A \left[(S_j^A - \langle S^A \rangle)-({E}_j^A - \langle {E}^A\rangle)\frac{C_3}{C_1}\right] 
\label{AEqMCoF}
\end{align}
Here, $t$ and $\tau$ have been replaced with a dimensionless time $t^* \equiv \int_0^t \frac{1}{\tau(\vec{p}(t'))}dt'$. Finally, subsystem $B$ can be assumed to be a thermal reservoir by assigning it an arbitrarily large energy such that it permanently remains in a state of stable equilibrium. As a result, it does not require an equation of motion and simply serves as a source or sink of the energy exchanged with the evolving subsystem $A$~\cite{Li2016a, Li2018}. This device is notated by recognizing that the ratio of co-factors in Equation~\ref{AEqMCoF} can be replaced by the thermodynamic factor, $\beta^R$, such that
\begin{align}
\beta^R = \frac{C_3}{C_1}= \frac{1}{k_b \, T^R}
\end{align}
where $T^R$ is the reservoir temperature.  

Equation~\ref{AEqMCoF} represents the SEAQT equation of motion for subsystem $A$, which consists of PEI-MP in a polar solvent (water) and a rare earth ion (Eu$^+3$). Subsystem $A$ can receive energy in a heat interaction from a thermal reservoir or it can expel energy in a heat interaction to that reservoir. In subsequent expressions, the  probabilities $p_j$, the energy $E$, and the entropy $S$ are understood to be those of subsystem $A$ so that the $A$ superscripts are omitted. Equation~\ref{AEqMCoF} is a system of 5800 coupled, first-order differential equations in time (one for each of the allowable energy levels). The system of equations can be solved numerically to yield the time-dependence of each of the occupation probabilities, $p_j$. These probabilities  can be combined through Equations~\ref{ExpectationE} to \ref{ExpectationValues}  with the eigenenergies, $E_j$, of the energy eigenstructure and their degeneracies $g_j$ (Figure~\ref{DOS}) to determine how all of the system properties evolve over time from any arbitrary initial state to stable equilibrium.  When the sequestration system, subsystem $A$, is fully equilibrated with the reservoir, the probabilities $p_j$ of Equation~\ref{AEqMCoF} become the same as the canonical stable equilibrium probabilities such that
\begin{align}
p_j^{{se}}=\frac {g_j \exp(-\beta^R \, E_j)}{\underset {j} {{\sum}}\phantom{l}g_j \exp(-\beta^R \, E_j)}
\label{CanDist}
\end{align}

Applying the equation of motion to identify the thermodynamic path involves establishing the energy eigenstructure (energy levels and degeneracies, i.e., Figure~\ref{DOS}) and then solving Equation~\ref{AEqMCoF} through the following steps:
\begin{enumerate}
\item Select an initial state by specifying a set of initial occupation probabilities, $p_j^{\,\tiny \text{init}}$. This set of probabilities establishes the entropy and energy, $\langle S \rangle = S^{\,\tiny \text{init}}$ and $\langle E \rangle = E^{\,\tiny \text{init}}$, respectively, of the initial state via Equations~\ref{ExpectationS} and (\ref{ExpectationE}.
\item Select a temperature for the thermal reservoir. The reservoir temperature establishes the canonical probabilities for the stable equilibrium state (the eventual state to which the equation of motion, Equation~\ref{AEqMCoF}, converges). The final stable equilibrium probabilities also can be found from Equation~\ref{CanDist}, and they determine the final entropy and energy, $\langle S \rangle =S^{\,\tiny \text{ final}}$ and $\langle E \rangle = E^{\,\tiny \text{ final}}$, respectively, via Equations~\ref{ExpectationS}  and \ref{ExpectationE}.
\item Solve Equations~\ref{AEqMCoF} numerically to obtain the unique kinetic path that maximizes entropy production from the initial state, $(S^{\,\tiny \text{init}},E^{\,\tiny \text{init}})$, to the final state, $(S^{\,\tiny \text{final}},E^{\,\tiny \text{final}})$.  Energy in a heat interaction can flow in either direction between the reservoir and the sequestration subsystem, $A$, depending upon the initial state and the reservoir temperature.
\end{enumerate}

The probability distribution of the initial state, Step 1, can either be formulated as a stable equilibrium state at a temperature different from the selected reservoir temperature using a canonical distribution (Equation~\ref{CanDist}), or it can be formulated as a non-equilibrium state using the procedure of Li and von Spakovsky \cite{Li2016a, Li2016b}. The later entails finding a partially canonical distribution $p_j^{{pe}}$ of the form

\begin{align}
p_j^{{pe}}=\frac {{\delta}_j g_j \exp(-\beta^{\text{pe}} E_j)}{\underset {j} {{\sum}}\phantom{l}{\delta}_j g_j \exp(-\beta^{\text{pe}} E_j)}
\label{PartCanDist}
\end{align}
and then perturbing it. This set of partial equilibrium probabilities has a form similar to that of the stable equilibrium set except for the inclusion of the factor, ${\delta}_j$. This factor is chosen to be either 0 or 1 for each energy level and is used to populate or depopulate specific energy levels. This partially canonical set of probabilities is then perturbed using the following expression:
\begin{align}
p_j^{\,\tiny \text{init}}= \lambda \; p_j^{{pe}} + (1-\lambda)\; p_j^{{se}}
\label{perturbed}
\end{align}
where $\lambda$ is chosen as a value between 0 and 1. This procedure provides a systematic way of conveniently controlling how close to or far away from stable equilibrium the initial non-equilibrium state is.

\subsection{Non-equilibrium Thermodynamic Paths}\label{SectionIID}

The SEAQT equation of motion was applied to the energy eigenstructure of Figure~\ref{DOS} with four different sets of initial states and thermal reservoir temperatures to explore a variety of kinetic behaviors. Solving the equation of motion for each combination of initial condition and reservoir temperature leads to a different kinetic path to equilibrium. In the following sections, the paths are labeled by Roman numerals I, II, III, IV (corresponding to red, blue, orange, and green curves, respectively, in the accompanying figures). Each path starts at an initial state indicated by an ``$i$'' subscript (e.g., IV$_\text{i}$) and ends at a final equilibrium state indicated by an ``$f$'' subscript (e.g.,  IV$_\text{f}$). Intermediate steps along a kinetic path that correspond to increasing times are indicated by numerical subscripts (e.g., IV$_2$, IV$_3$, IV$_4$, and IV$_5$). 

\begin{figure}[ht!]
	\begin{center}
		\includegraphics[width=0.50\textwidth]{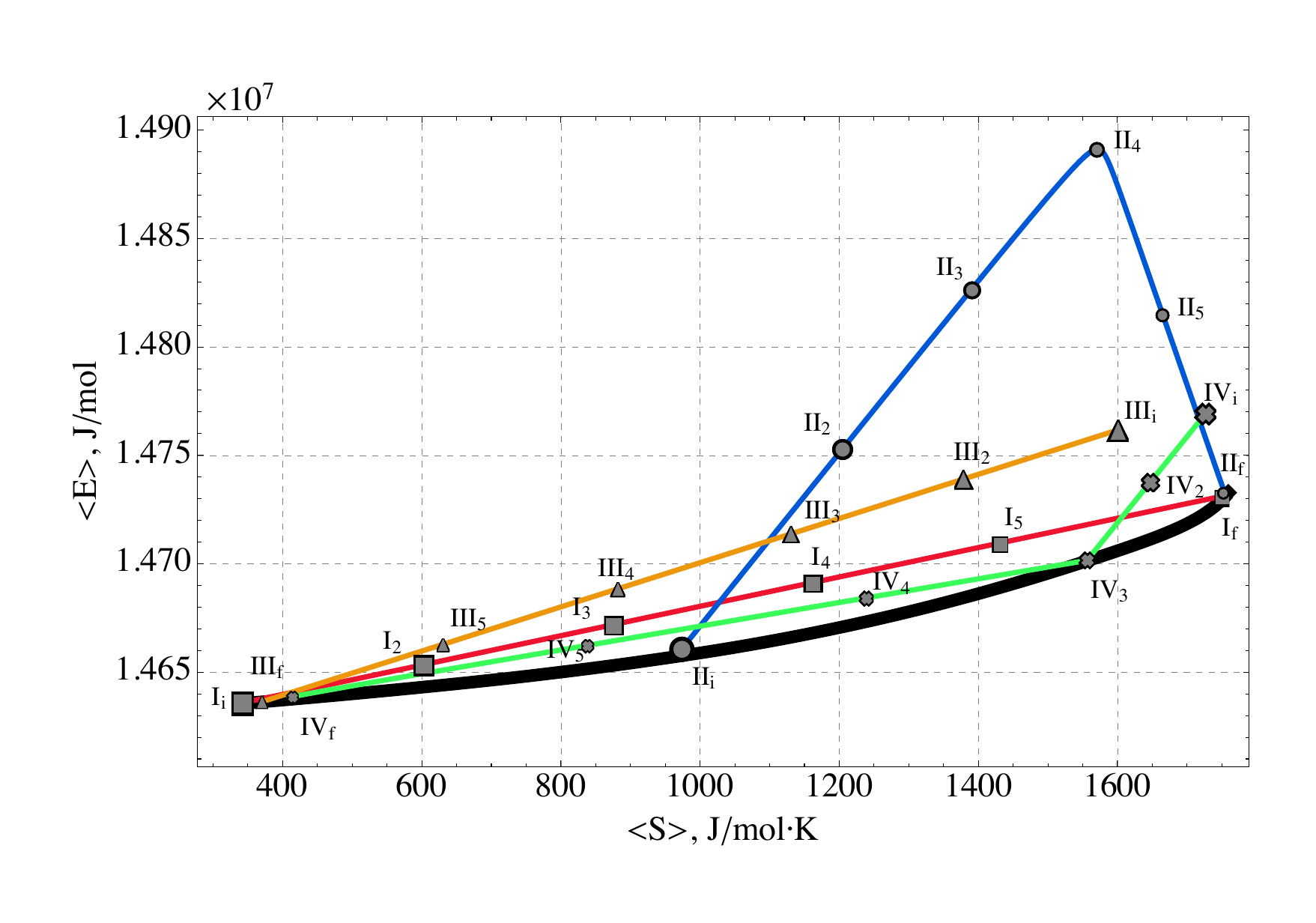}
		\caption{Four non-equilibrium thermodynamic paths to stable equilibrium. Path I (red) represents heating from an initial low-temperature stable equilibrium state to a state in mutual stable equilibrium with a high-temperature reservoir at 350 K; Path II (blue) involves heating then cooling from an intermediate-energy non-equilibrium state to mutual stable equilibrium with the same high-temperature reservoir as is present for Path I; Paths III and IV involve cooling from high-energy non-equilibrium states to mutual stable equilibrium with a low-temperature reservoir at 5 K.  The labeled points along each path indicate the direction of the paths from the initial state ({\em i\/}) to the final ({\em f\/}) state.}
		\label{Thermodynamic_Paths}
	\end{center}
\end{figure}

The four paths are indicated by the labeled and colored curves on the $E$ vs $S$ diagram of Figure~\ref{Thermodynamic_Paths}. The energy and entropy units are reported in terms of J/mol and J/mol K, respectively, where the energy indexes from the Wang-Landau results are first scaled to their original magnitudes (reversing any scaling that was done to adjust the precision of the energy levels). Next, the results are multiplied by Boltzmann's constant $k_B$ and then divided by the number of atoms in the simulated system, i.e., 3285, and multiplied by Avogadro's number, thus, arriving at units of J/mol. The heavy black curve in Figure~\ref{Thermodynamic_Paths} represents the loci of stable equilibrium states. Any state above and to the left of the black curve is not a stable equilibrium state (e.g., it is a non-equilibrium state or a partially canonical state). It is important to remember that temperature is the tangent to the loci of stable equilibrium states, that is, $T \equiv (\partial{E}/\partial{S})_{V,n}$ where $E$, $S$, and $V$ are expectation values of extensive quantities and $n$ is the fixed number of molecules in the system. Temperature is only defined along the loci of stable equilibrium states. It does not have meaning for non-equilibrium states because the partial derivative is undefined. Also, for simplicity, we only include bonding energy in the energy eigenstructure. Thus, the temperature here only represents a notional temperature rather than the actual temperature. The latter would require including the usual electronic and vibrational contributions to heat capacity in Equation~\ref{eqn:Ej} and the energy eigenstructure of Figure~\ref{DOS}.

Solving the SEAQT equation of motion yields the time-dependence of the occupation probability for each energy level. Once it is established how these probabilities evolve with time, they can be used to determine the evolution of the enrtopy production and  thermodynamic properties like the energy and entropy. The time steps of the SEAQT equation of motion move through state space rather than physical space. Consequently, the relaxation parameter $\tau$ in Equation~\ref{ABEqM} (or implicit in the non-dimensional time, $t^*$, of Equation~\ref{AEqMCoF}) must be fitted to experimental data or estimated from semi-empirical theoretical models of specific kinetic mechanisms in order to connect the SEAQT kinetics to real times. Details of how $\tau$ is fitted to real time are provided in Section~\nameref{Discussion}.

It should be noted here that the scaling provided by the time relaxation parameter is a common explicit or implicit feature of dynamic models at all levels of description. For example, the mass, momentum, and energy balances of continuum models require experimentally determined transport coefficients, which scale the predictions made by these balances to real time. This is also true for mesoscopic models where the time relaxation parameters appearing in the Boltzmann transport equations, the BGK equation, the Fokker-Planck equation (implicitly via its drift and diffusion coefficients), etc. are based on either experimental or semi-empirical values that place the predictions made by these equations in real time. In fact, Li, von Spakovsky, and Hin \cite{Li2018steepest} have shown that the Boltzmann transport equations are a special case of the SEAQT equation of motion. Furthermore, as shown by Worden, von Spakovsky, and Hin \cite{worden2023}, using experimentally determined characteristic electron and phonon relaxation times specific to a given semiconductor material, the thermal and electrical properties of these materials across a wide range of temperatures can be accurately predicted with the SEAQT equation of motion. It is important to keep in mind that the choice of relaxation parameter does not affect  predicted property trends. It simply scales the trends to real time. This is also true of the quantum Lindblad equation that requires experimental values for its relaxation parameters. The Schrödinger equation of motion, of course, has no need of relaxation parameters, but then its predictions are not in real time since its motion is that of a reversible process and is, thus, unable to correctly predict the behavior of a quantum computer due to the presence of the dephasing and relaxation phenomena. In contrast, both the Lindblad and SEAQT equations can \cite{jhon2020,jhon2022,cano2015steepest}, provided they have relaxation parameter values scaled to experimentally determined times specific to the dephasing and relaxation characteristics of a given quantum computer. Of course, the relaxation parameter does make the SEAQT equation of motion`s dissipation {\em term} phenomenological in character but this is the case as well for the other equations mentioned. Nonetheless, the dissipation {\em operator} itself, is not phenomenological since it is based upon a quantum thermodynamic non-linear dynamic in state space specific to the steepest-entropy-ascent principle that fundamentally captures the quantum or classical dynamics of a given system. 


Finally, the four non-equilibrium paths generated in this work are shown in Figure~\ref{Thermodynamic_Paths}; they reflect heating and cooling under the following kinetic conditions:

\begin{description}
\item[Path I (red)] describes gradual heating from a low initial temperature of 5 K (notional temperature). The initial energy, $\langle E \rangle =1.464 \times 10^{7} \frac{\text{J}}{\text{mol}}$, corresponds to stable equilibrium at this temperature. From this initial state, the reservoir is set to a higher temperature (350 K) that corresponds to a final equilibrium energy of $\langle E \rangle =1.473 \times 10^{7} \frac{\text{J}}{\text{mol}}$. 

\item[Path II (blue)]
reflects initial heating  with the same reservoir temperature as for Path I followed by cooling, but from an initial non-equilibrium state. Steepest entropy ascent from the non-equilibrium starting point selects a path that increases the energy more rapidly than along Path I and the energy of the sequestration subsystem ``overshoots'' before decreasing to the final value corresponding to equilibrium with a reservoir at 350 K (same final state as for Path I).

\item[Path III (orange)]
corresponds to cooling from an initial non-equilibrium state with a relatively high energy. The final equilibrium state is the same as the initial state of Path I; the reservoir is at 5 K.

\item[Path IV (green)]
corresponds to cooling from an initial non-equilibrium state that has a higher energy and a higher entropy than those of the initial state of Path III. The final equilibrium state is the same as the initial state of Path I; the reservoir is at 5 K.
\end{description}

\subsection{Linking State-Based 
Evolution to Microstructural Evolution}\label{SectionIIE}

Before describing the evolution of energy and entropy along the four paths in state space, we describe how to link the thermodynamic states to specific physical characteristics of the system. Previous work applying SEAQT to energy eigenstructures generated with the Replica Exchange Wang-Landau algorithm has shown that paths in state space can be linked to physical representations by generating expected values of properties or descriptors from the occupation probabilities along the kinetic path~\cite{McDonald2022,Younis2022,2023MvSR_arXiv_brush}. However, because most systems larger than a few molecules have astronomically large degeneracies, storing even a small fraction of the properties or conformations associated with each energy level is impractical.  Nonetheless, the problem can be made tractable by recording a set of descriptors (parameters that can be calculated from each recorded configuration sampled by Wang-Landau) when the energy eigenstructure is first generated. It turns out that most energy levels along a particular kinetic path have essentially zero probability of being occupied, i.e., the energy levels with a fairly significant non-zero occupation probability are few in number. Once the non-zero energy levels are identified by the SEAQT equation of motion, the Replica Exchange Wang-Landau algorithm can be run a second time with a flag that records representative descriptors and  configurations for only these occupied energy levels, which form a small subset of all possible energy levels. This procedure for linking state space to microstructure is described in detail in reference~\cite{McDonald2022}~. The three primary descriptors chosen for the sequestration system are the neighbor distribution of the charged elements, $n_d(r)$, the radius of gyration, $R_g$, and the tortuosity, $\zeta$~\cite{McDonald2023polymer}. Some secondary descriptors, as described below, are also used.

It is worth noting that the Replica Exchange Wang-Landau algorithm is more effective at sampling the domain of possible conformations than traditional Metropolis Monte Carlo molecular modeling methods because the Replica Exchange method moves relatively freely through a system's energy space whereas the Metropolis approach has thermal constraints on possible conformational changes.

\subsubsection{System Descriptors}

As mentioned above, the three primary parameters used to characterize the system include the radius of gyration of the polymer chain and functional groups, the tortuosity of the PEI backbones, and charged-element neighbor distributions. The radius of gyration reflects how the mass of the chain is distributed and is determined from the expression,

\begin{equation}
R_g = {\left(\frac{1}{N} \sum_i^N   \lVert \boldsymbol{r}_i-\boldsymbol{r}_{\textrm{cm}}\rVert{}^2\right)}^{\frac{1}{2}}
\end{equation}
Here, $R_g$ is dependent upon the number of monomers in the chain, $N$. The quantity, $r_i$, represents the 3D coordinates of a single monomer, and $r_{cm}$ are the 3D coordinates of the center of mass. All the PEI monomers, including the functional groups, are assumed to have the same mass. 

The tortuosity reflects the extent of bending along the polymer chains and is found from
\begin{equation}
\zeta ={\left(\frac{1}{N-2} \sum _i^{N-2} \lVert\boldsymbol{s}_i-\bar{\boldsymbol{s}}\rVert{}^2\right)}^{\frac{1}{2}}
\end{equation}
where $\boldsymbol{s}_i$ is obtained from
\begin{equation}
\boldsymbol{s}_i= \left( {\begin{array}{cc}
   {\sum_{j_{\boldsymbol{x}}=1}^i w_{j_{\boldsymbol{x}}}} \\
   {\sum_{j_{\boldsymbol{y}}=1}^i w_{j_{\boldsymbol{y}}}} \\
   {\sum_{j_{\boldsymbol{z}}=1}^i w_{j_{\boldsymbol{z}}}} \\
  \end{array} } \right)
  \end{equation}
with
  	\begin{equation}
	\boldsymbol{w}_j=\boldsymbol{r}_{j,j+1}\times \boldsymbol{r}_{j,j+2} , \; 1 \leq i \leq (N-2)
	\end{equation}
The tortuosity is a function of the number of turns along the chain's length. The calculation is performed by taking the cross-product of the coplanar vectors of two successive monomers. The values for each vector component of $w_j$ are then summed for all previously checked monomers in the sequence and assigned to $\boldsymbol{s}_i$, where $i$ denotes the currently inspected monomer.  The tortuosity calculated here is 3-dimensional, and because variable bond lengths are allowed in the model, $\boldsymbol{s}_i$ has more than one non-zero component~\cite{Wust2012} and the computed components do not necessarily equal $1$ or $-1$.

The neighbor distribution function, $n_d(r)$, is defined to characterize the distribution of ions relative to charged binding sites, i.e.,
\begin{equation}
n_d(r)=\frac{\rho_l(r)}{\rho_i}
\label{NDF_equation}
\end{equation}
where ${\rho_l}(r)$ is a function of the local density of a component at a given neighbor distance that depends upon the number of accessible cells of the lattice at each discrete distance, and $\rho_i$ is the density over the whole system of the paired $i$ constituent. The calculation for the neighbor distribution is similar to the radial distribution functions reported in many molecular dynamics simulations~\cite{Archer2022Resins}. 

As to the secondary parameters (descriptors) used to characterize the sequestration process, they include the average number of sequestered ions per energy level and the average number of functional groups per polymer ion complex. Functional complexation with ions is assumed to occur when the number of nearby attractive charges exceeds 5 and the ion is attracted by at least 2 functional groups. The interactions considered for these parameters are limited to distances less than $a\sqrt{10}$ to mimic a first coordination shell for the complexes. This reduces the over-counting of interactions among constituents. 

For each descriptor, a single arithmetic average value of the recorded configurations for each energy level is stored for that level of the energy eigenstructure. An expectation value for each descriptor at each instant of time is obtained from expressions analogous to those of Equations~\ref{ExpectationE} to \ref{ExpectationValues}, using these arithmetic averages and the probabilities  predicted at each instant of time by the SEAQT equation of motion, Equation~\ref{AEqMCoF}. 

\section{Results}\label{SectionIII}

\subsection{Thermodynamic Properties}

\begin{figure}
	\centering
	\includegraphics[width=.47\textwidth]{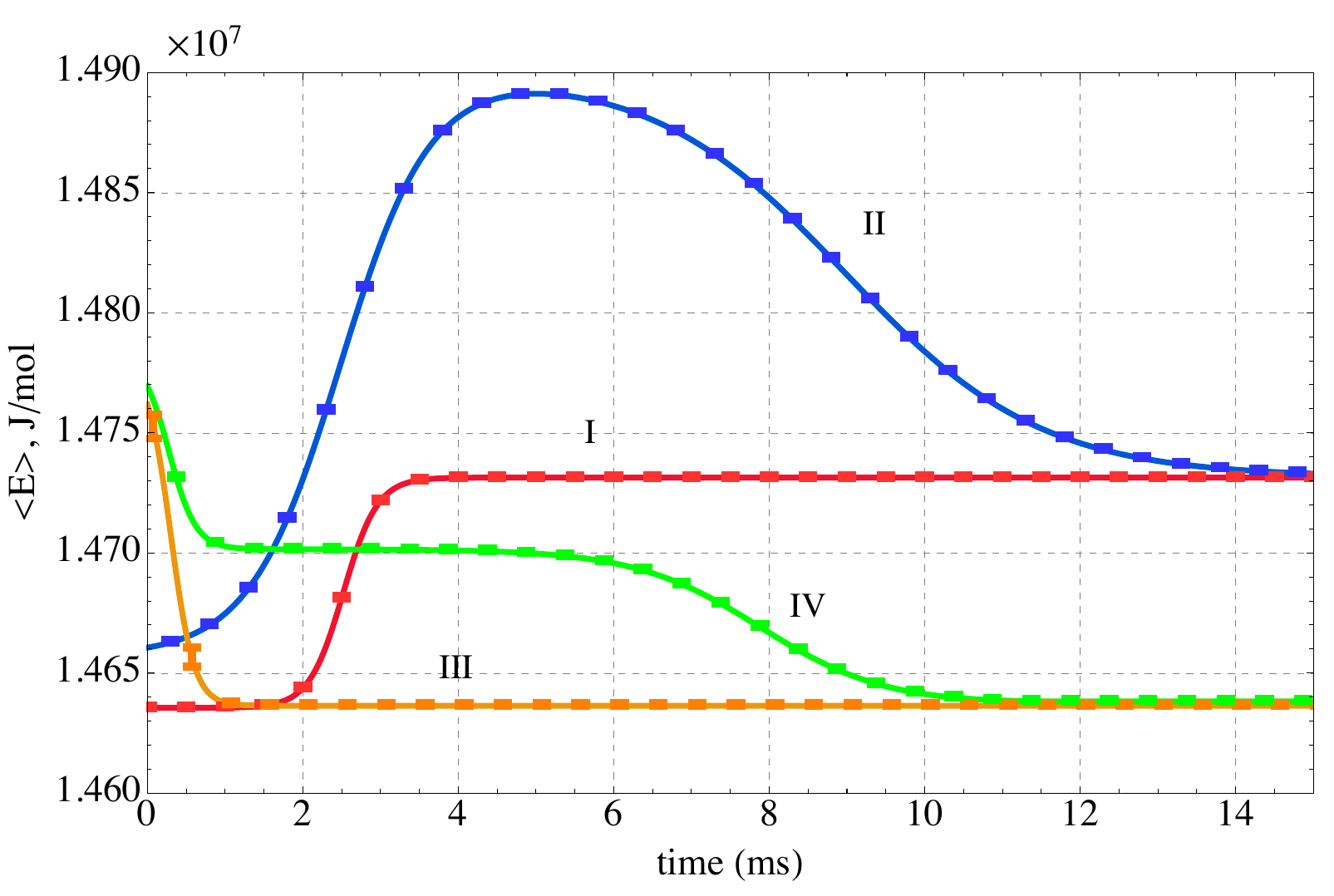}
	\caption{ The time evolution of the energy of the sequestration subsystem ($A$) along the four kinetic paths defined in Figure~\ref{Thermodynamic_Paths}. The sequestration system is heated along Path I, heated and cooled along Path II, and cooled along Paths III and IV.}
	\label{EP_1}
\end{figure}

The evolution of state of the sequestration subsystem (subsystem $A$) composed of three polymer chains in a polar solvent with rare earth ions exchanging energy with a thermal reservoir (subsystem $B$) is predicted by the SEAQT equation of motion. The time evolution of the sequestration subsystem's energy along each of the four kinetic paths is shown in Figure~\ref{EP_1}.  The error bars along these paths in Figure~\ref{EP_1} and subsequent figures come from the maximum and minimum values of the error bars on the Wang-Landau density of states shown in Figure~\ref{DOS}.  The energy initially increases with time along Paths I and II as subsystem $A$ absorbs energy during heating. Path I, which starts in a canonical state, moves directly to mutual stable equilibrium with the high-temperature reservoir at 350 K. For Path II, the energy of the sequestration subsystem increases rapidly to a maximum and then decreases to the equilibrium energy established by the temperature of the high-temperature reservoir. The radically different energy trajectories for Paths I and II arise because their different initial states lead to different directions of maximum entropy production. The sequestration subsystem energy decreases along both Paths III and IV because energy is extracted from the sequestration subsystem during cooling.

\begin{figure}
	\centering
	\includegraphics[width=.47\textwidth]{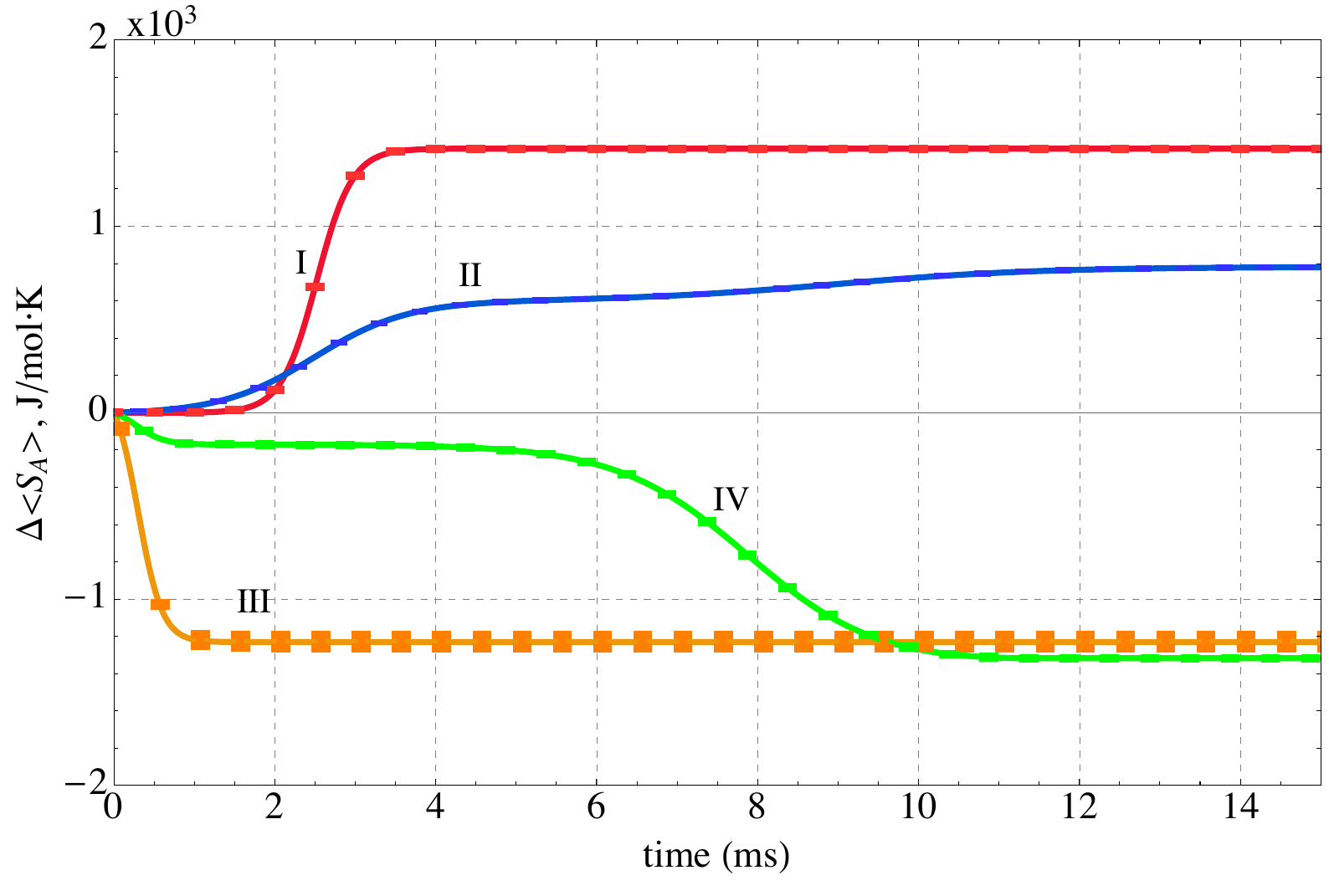}
	\caption{The time evolution of the entropy change in the sequestration subsystem $A$ along the four kinetic paths defined in Figure~\ref{Thermodynamic_Paths}. The sequestration system is heated along Path I, heated and cooled along Path II, and cooled along Paths III and IV.}
	\label{QP_1}
\end{figure}

A second way of following the kinetic progress along the paths is by plotting the change in entropy of the sequestration subsystem, $\Delta \langle S_{A} \rangle$, with time. This is done in Figure~\ref{QP_1}. The change in this entropy is affected by the direction of energy flow to or from the reservoir.  Thus, $\Delta \langle S_{A} \rangle$ increases along Path I because the sequestration subsystem generates entropy and receives entropy via a heat interaction with the high-temperature reservoir. Along Path II, it also increases during the heating phase for the same reason, while during the cooling phase the net effect of the entropy loss from the system and the entropy generated is positive. For Paths III and IV the entropy decreases since the entropy transfer via a heat interaction to the low-temperature reservoir is greater than the entropy generated by the sequestration subsystem.

The entropy produced during sequestration provides another representation of the process. The entropy change shown in Figure~\ref{QP_1}, $\Delta \langle S_{A} \rangle$, has two contributions: (i) one from entropy exchanged in a heat interaction with the reservoir at the temperature of the reservoir and (ii) a second contribution that arises from the creation of entropy within the sequestration subsystem, i.e., the entropy production, $\sigma$, found from the following entropy balance on the subsystem: 
\begin{align}
\Delta \langle S_{A} \rangle = \frac{\Delta \langle E_{A} \rangle}{T^{R}} + \sigma
\label{EntropyProduction}
\end{align}
both $\Delta \langle S_{A} \rangle$ and $\Delta \langle E_{A} \rangle$ are readily determined from the time-dependent occupation probability distributions. The entropy production is a consequence of the energy redistribution in time among the energy levels within the sequestration subsystem.
\begin{figure}
	\centering
	\includegraphics[width=.47\textwidth]{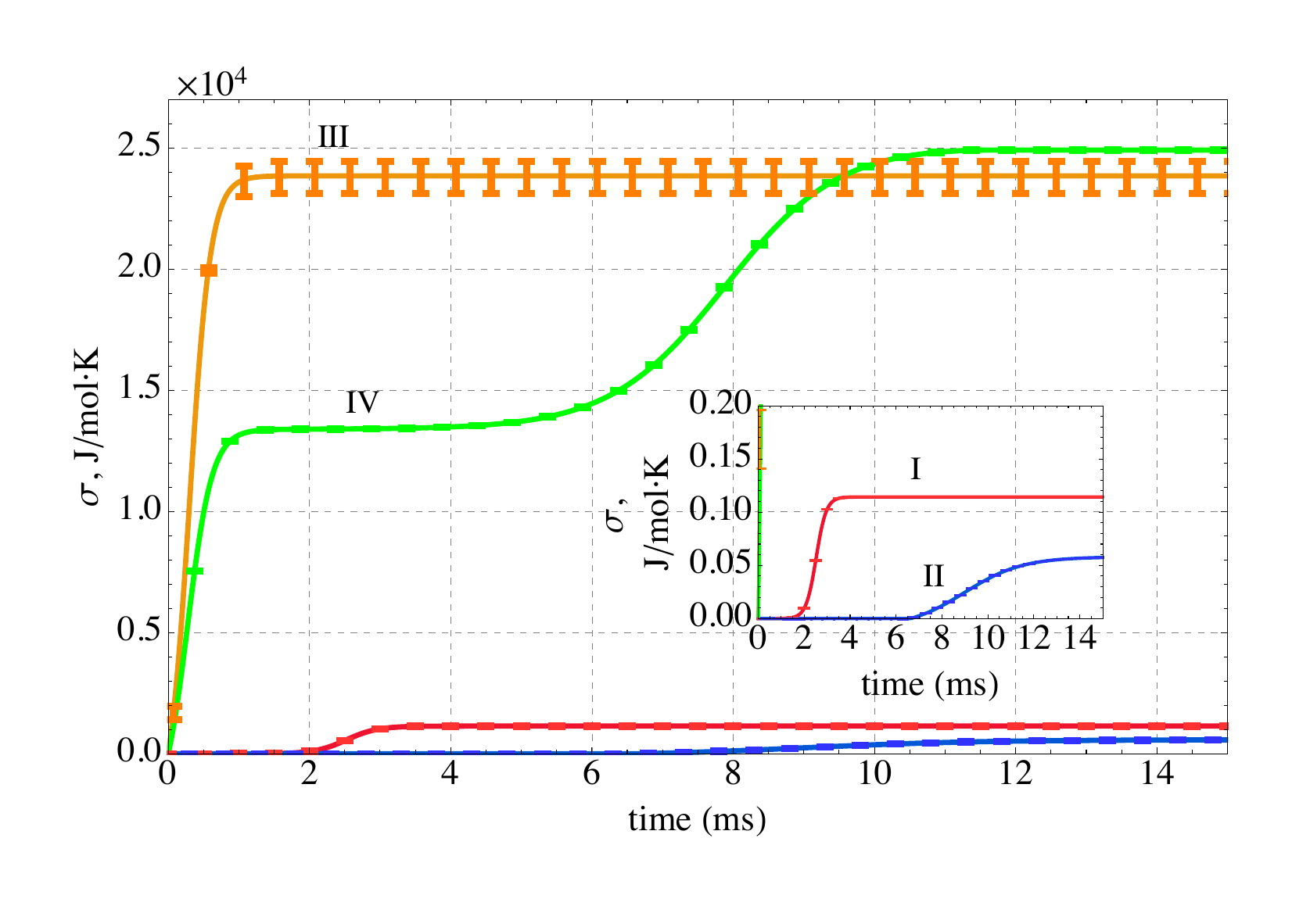}
	\caption{The time evolution of the entropy production along the four kinetic paths defined in Figure~\ref{Thermodynamic_Paths}. The sequestration system is heated along Path I, heated and cooled along Path II, and cooled along Paths III and IV. All the entropy produced is in the sequestration subsystem $A$. The inset enlarges the initial portion of Paths I and II for clarity.}
	\label{SP_1}
\end{figure}

The entropy production is shown for the four Paths in Figure~\ref{SP_1}.  Entropy production monotonically increases along all four paths.  The first two paths represent evolutions from initial states characterized by low-energy conformations of the polymer-ion complex and an unfolding of the multi-chains as the system is heated and a refolding in the case of Path II when it is cooled. The last two paths represent a coiling of the polymer chains as they cool and their energy decreases. Interestingly, of the four paths considered, the two cooling-only paths (III and IV) generate substantially more entropy than the two heating paths. The large increase in entropy during cooling happens when the system shifts from an energy with very high degeneracy to a lower energy with significantly lower degeneracy (see Figure~\ref{DOS}). Thus, the {\em direction} along a path in $E$-$S$ space (e.g., heating or cooling) has a large impact on the amount of entropy produced by the non-equilibrium process. Somewhat surprisingly, the proximity of the kinetic path to the loci of equilibrium states (the black curve in Figure~\ref{Thermodynamic_Paths}) has little to do with the entropy production.

\subsection{Physical Properties}

\begin{figure}
 	\centering
 	\includegraphics[width=.47\textwidth]{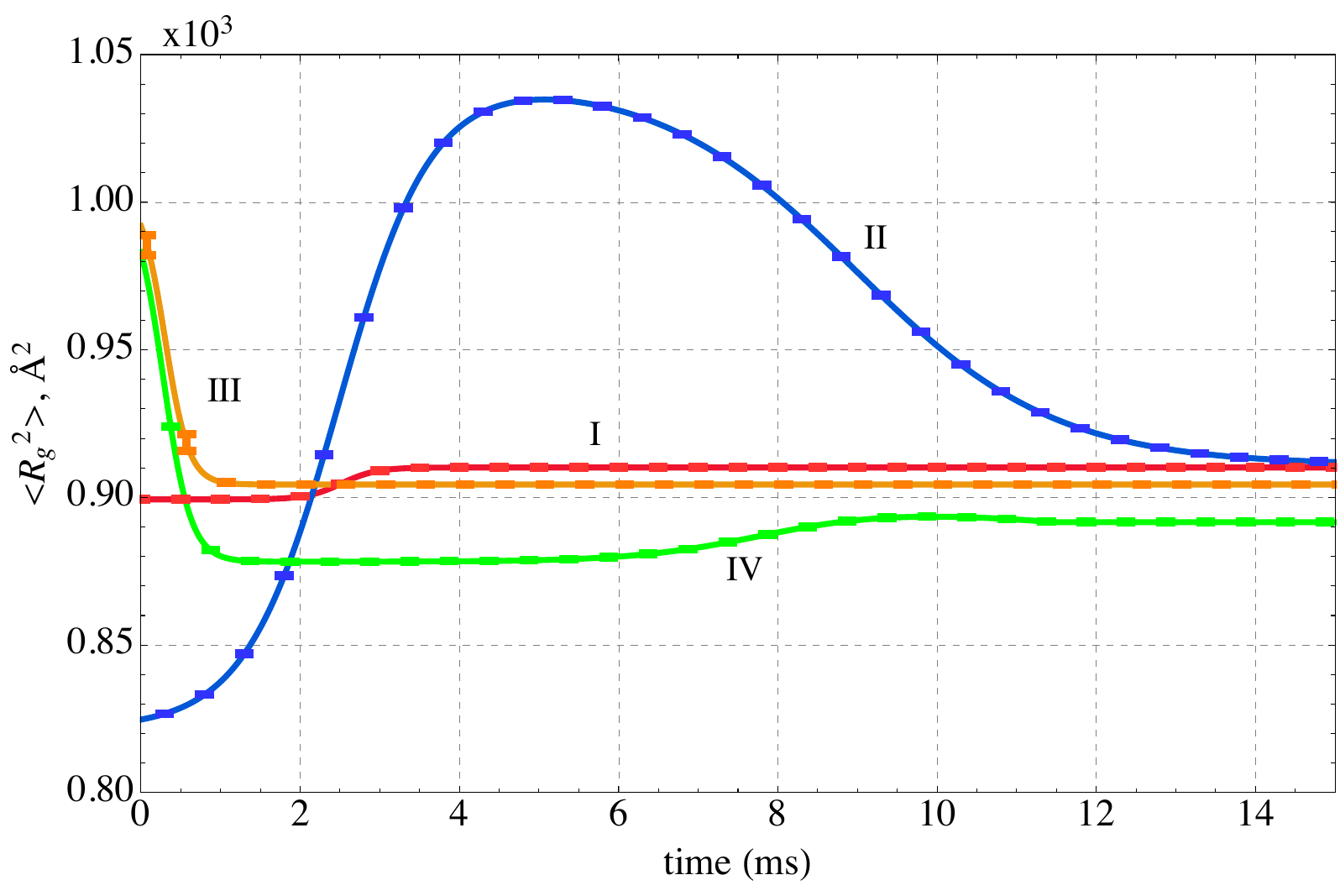}
 	\centering
 	\includegraphics[width=.47\textwidth]{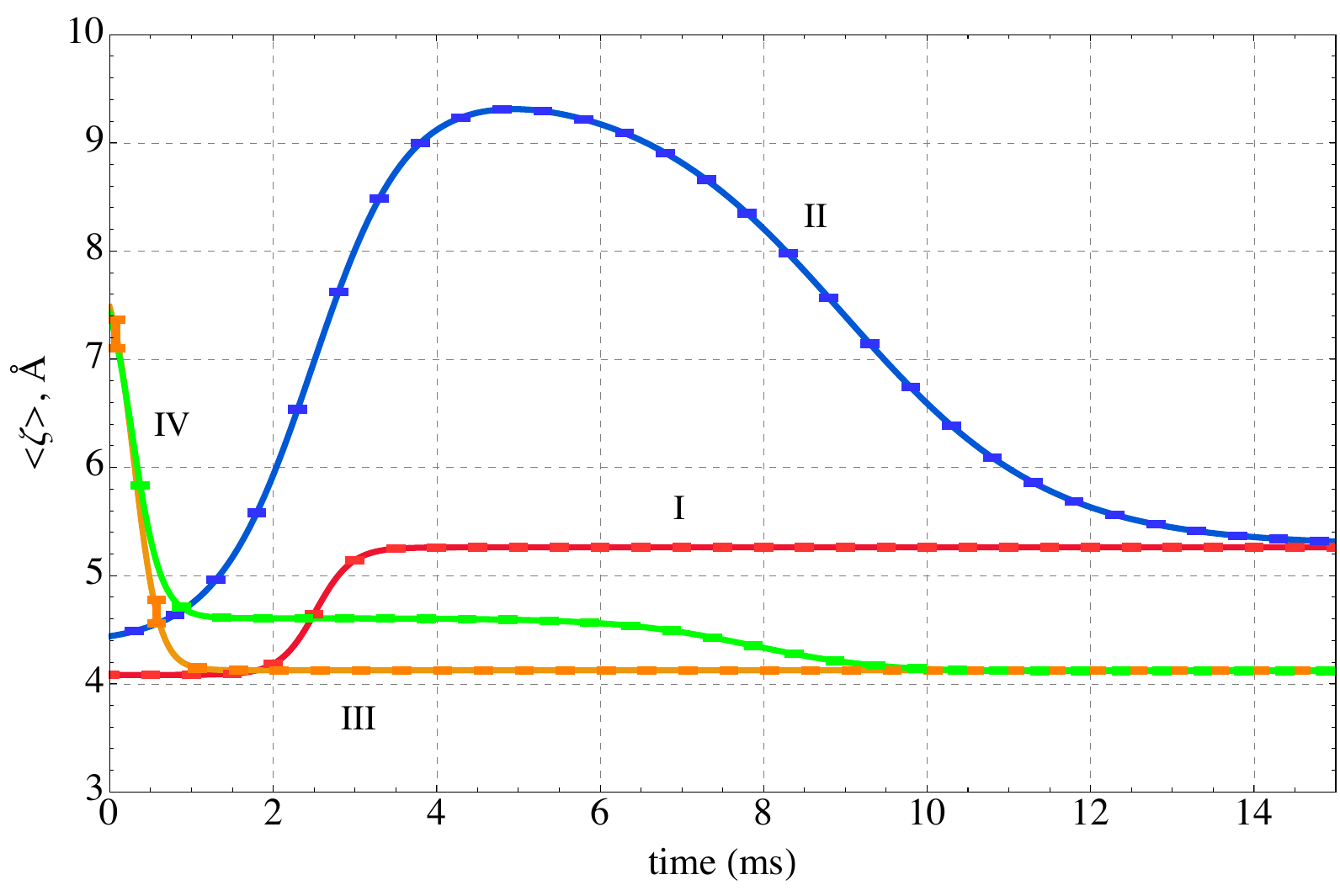}
 	\hspace*{\fill} 
 	\caption{Time evolution of the tortuosity and the square of the radius of gyration along the four non-equilibrium paths. The sequestration system is heated along Path I, heated and cooled along Path II, and cooled along paths III and IV.}
 	\label{Rg_t_1}
 	\hspace*{\fill} 
 \end{figure}

The evolution of the expectation values of the physical properties or descriptors are shown in this section.  For example, the time evolutions of the expectation values of the tortuosity, $\zeta$ and the square of the radius of gyration, $\langle R_{g}^2 \rangle $, are given in Figure~\ref{Rg_t_1}. The increases and decreases of these properties in the figure largely reflect the time-dependent changes in the system energy shown in Figure~\ref{EP_1}. Increases in $\langle R_{g}^2 \rangle $ represent expansion of the PEI-MP chain whereas decreases indicate the chain is contracting toward a compact configuration with lower energy. Conversely, increases in $\zeta$ correspond to a lengthening or straightening of the PEI-MP chain, and decreases in $\zeta$ reflect contraction toward a compact configuration.
 
\begin{figure}[ht!]
	\begin{center}
		\begin{subfigure}{0.45\textwidth}
			\includegraphics[width=\linewidth]{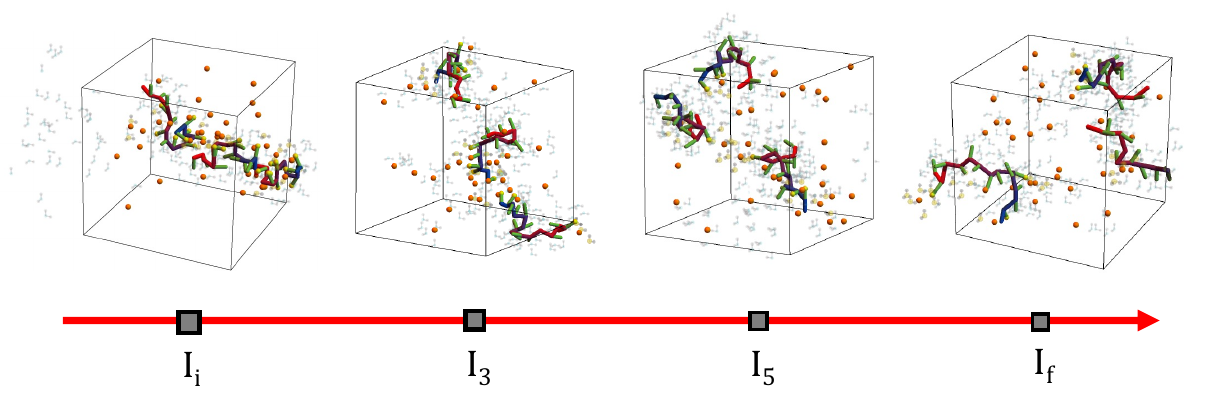}
			\caption{} \label{fig:conf-a}
			\includegraphics[width=\linewidth]{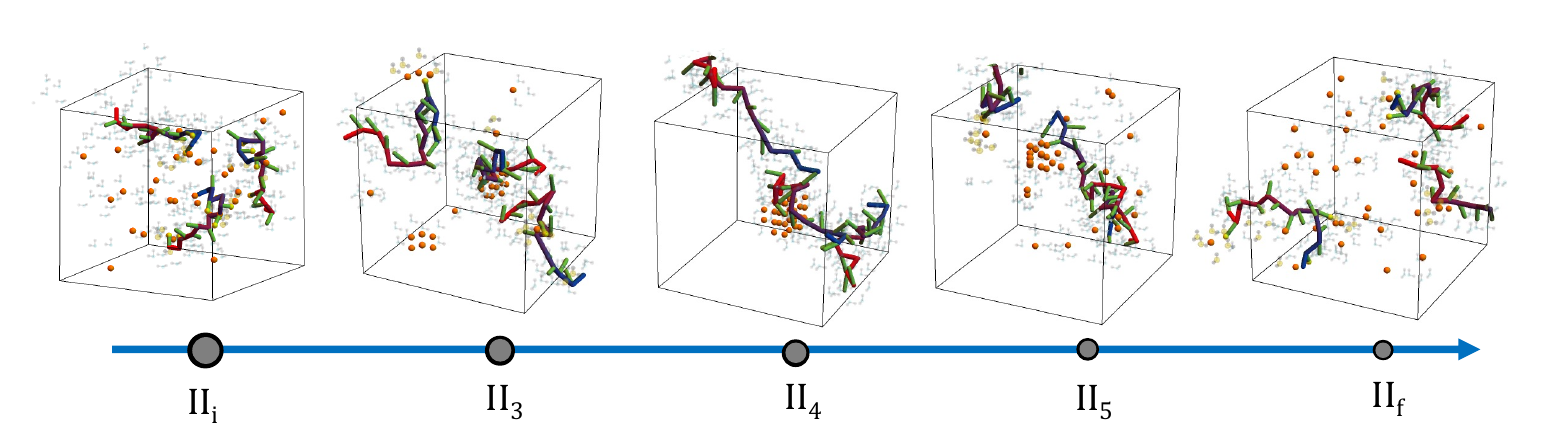}
			\caption{} \label{fig:conf-b}
			\includegraphics[width=\linewidth]{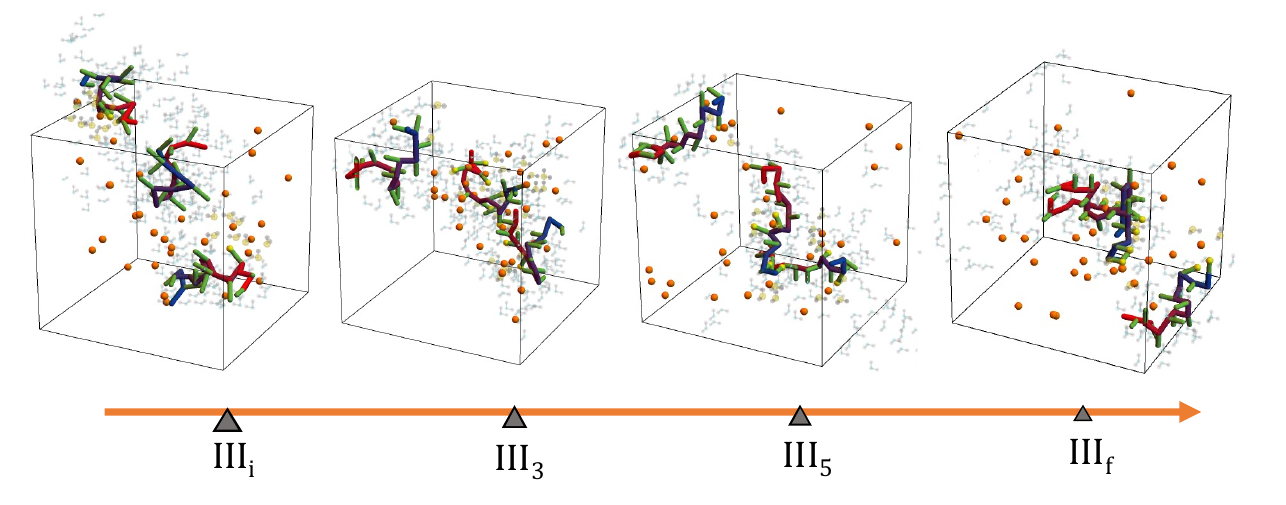}
			\caption{} \label{fig:conf-c}
			\includegraphics[width=\linewidth]{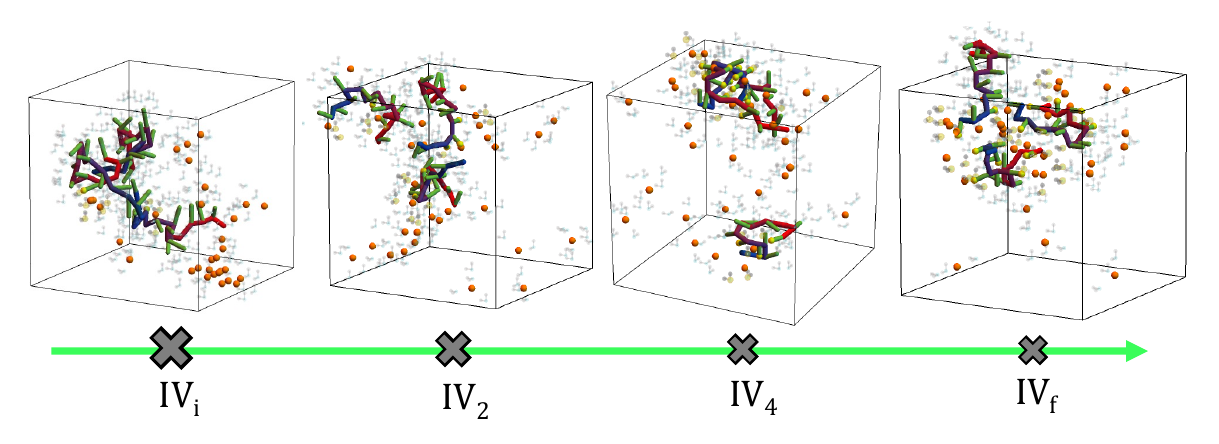}
			\caption{} \label{fig:conf-d}
		\end{subfigure}%
		\hspace*{\fill}
		\caption{Time evolution of the sequestration conformations along the four non-equilibrium paths. The system is heated along Path I, heated and cooled along Path II, and cooled along paths III and IV.}
		\label{RepMicro_1}
	\end{center}
\end{figure}

As with the physical properties or descriptors, the physical appearance of the sequestration system is also constructed from the evolving occupation probabilities predicted by the SEAQT equation of motion. Time sequences of the sequestration configurations along each of the four paths are shown in Figure~\ref{RepMicro_1}. Each  configuration in this figure is selected to match the $\langle E \rangle$, $\langle R_{g} \rangle$, and $\langle \tau \rangle$ values of the corresponding thermodynamic state.  Because the expected values for these properties are generated from more than $10^7$ states sampled by the first REWL simulation and there are approximately $10^6$ conformations visited by the second REWL simulation, each configuration in Figure~\ref{RepMicro_1} is selected using information from a large number of  individual conformations.  The first configuration of Path I (left side of Figure~\ref{RepMicro_1}(a)) has the lowest energy of this path, and the energy steadily increases until stable equilibrium is reached at the reservoir temperature (the configuration on the right).  Path II is unique in that the energy increases from its lowest value (initial state) to a maximum at II$_4$ and then decreases again to the stable equilibrium state on the right. The energy for Paths III and IV is highest for the initial state and decreases during cooling to a low value at stable equilibrium. The configurations reflect this change, showing a continual contraction of the configuration as the energy decreases.

\subsection{Sequestration Kinetics}

The molecular effects that influence the capture of Eu$^{+3}$ from solution include  
all the interactions among the charged species of the system. A rare-earth ion can be sequestered by a negatively charged specie, or a positively charged specie can repel it and reduce its chance of being captured. For example, Eu$^{+3}$ can be electrostatically bound by the O$^-$ end of a water molecule and by the O$^-$ portion of an MP functional group. Furthermore, it can be excluded from the O$^-$ portion of an MP functional group by the H$^+$ portions of the PEI chain. All these interactions take place concurrently. They are characterized with neighbor distribution functions calculated from Equation~\ref{NDF_equation} for each pair of charged species. 

\begin{figure}[ht!]
	\begin{center}
	   \begin{subfigure}{0.45\textwidth}
    \includegraphics[width=\linewidth]{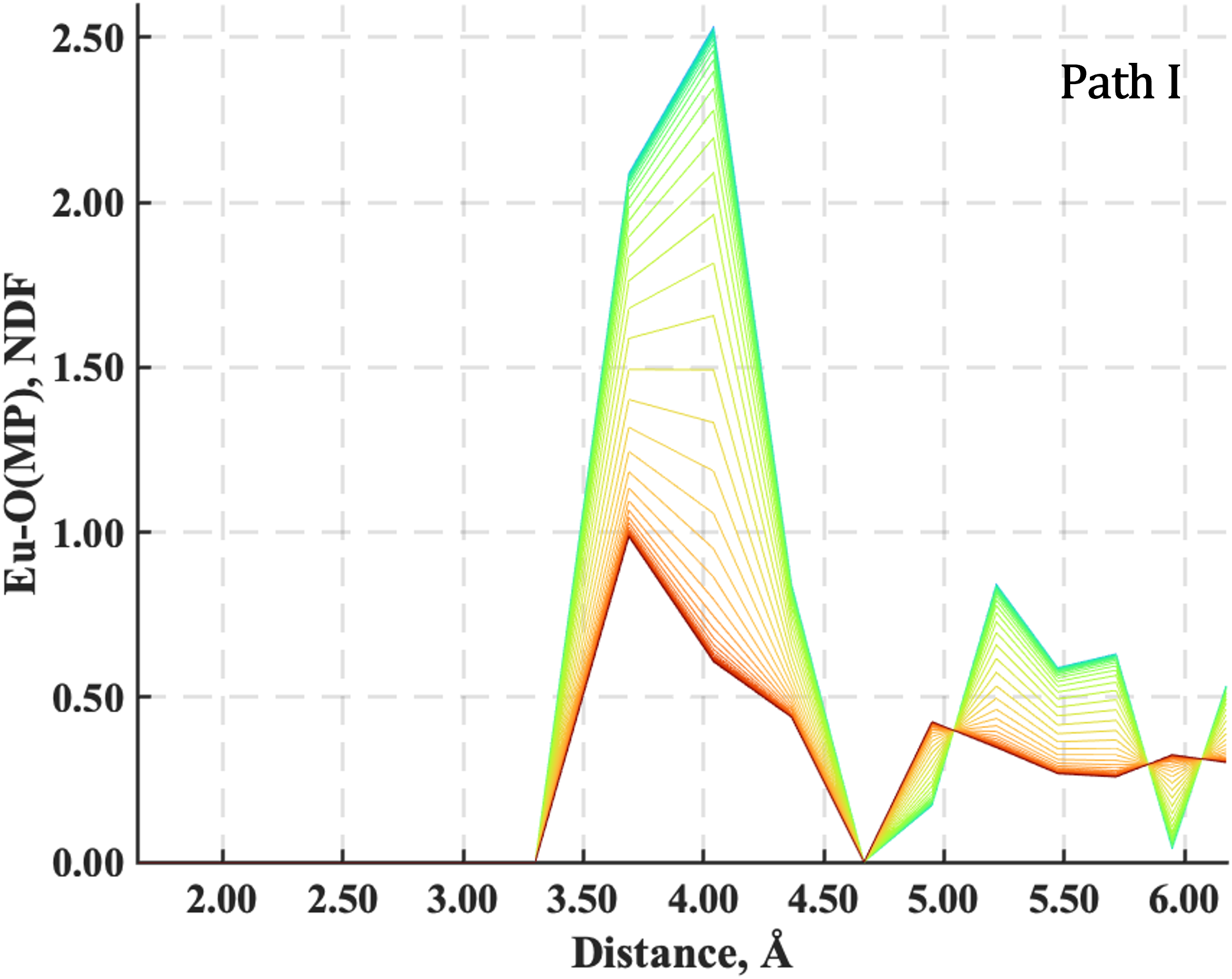}
    \caption{} \label{fig:NDF-I}
    \includegraphics[width=\linewidth]{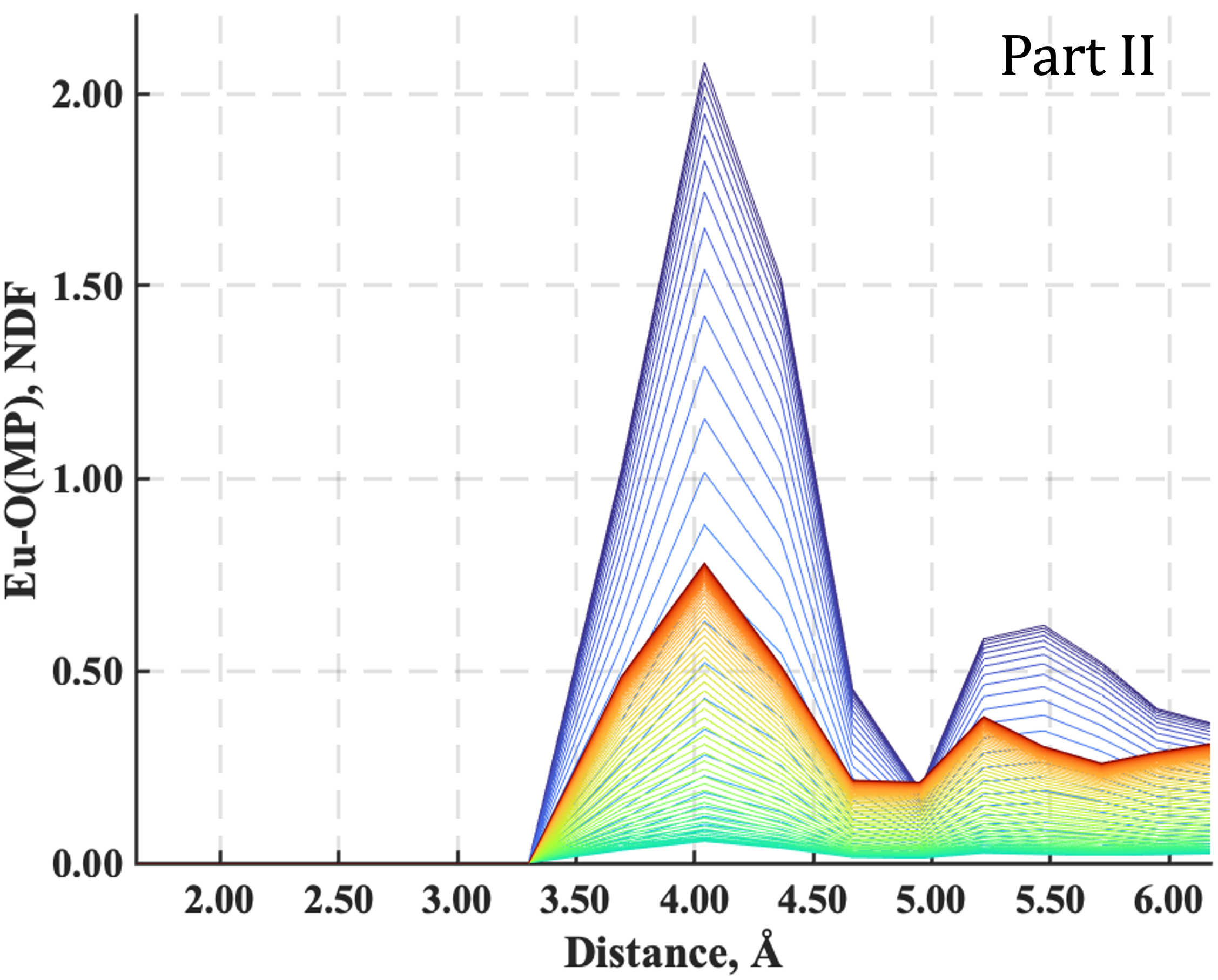}
    \caption{} \label{fig:NDF-II}
      \end{subfigure}%
  \hspace*{\fill}
		\caption{
  Neighbor distribution functions between O$^-$ on the MP functional group and Eu$^{+3}$ along Paths I and II. The family of colored curves represent different times along the respective paths from the initial state (blue) to stable equilibrium (red).}
  \label{NDFheating}
  \end{center}
\end{figure}

\begin{figure}[ht!]
	\begin{center}
	\begin{subfigure}{0.45\textwidth}
    \includegraphics[width=\linewidth]{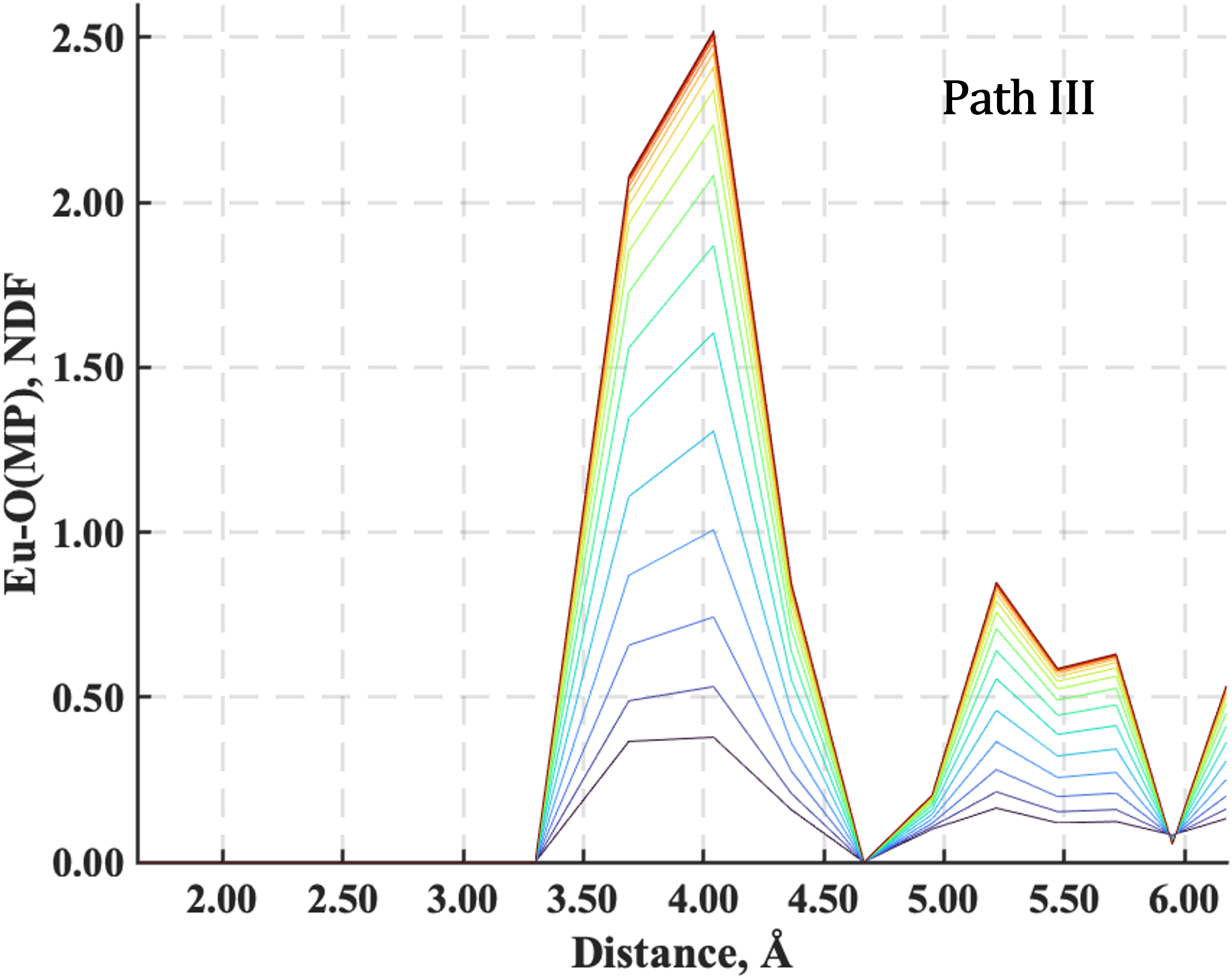}
    \caption{} \label{fig:NDF-III}
    \includegraphics[width=\linewidth]{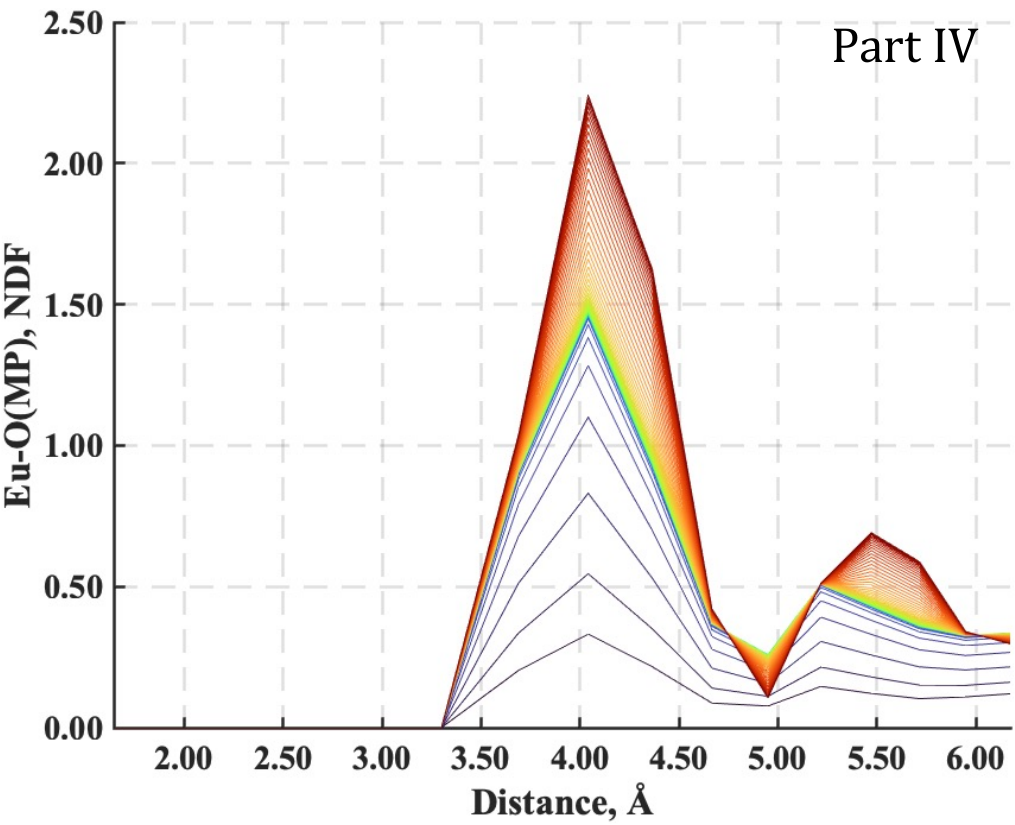}
    \caption{} \label{fig:NDF-IV}
      \end{subfigure}%
  \hspace*{\fill}
		\caption{
  Neighbor distribution functions between O$^-$ on the MP functional group and Eu$^{+3}$ along Paths III and IV. The family of colored curves represent different times along the respective paths from the initial state (blue) to stable equilibrium (red).}
  \label{NDFcooling}
  \end{center}
\end{figure}

The neighbor distribution function most sensitive to the extent of Eu$^{+3}$ sequestration is the one between the O$^-$ portion of an MP functional group and Eu$^{+3}$.  The variation of the neighbor distribution function for this pair, denoted as Eu-O(MP), is plotted in Figures~\ref{NDFheating} and \ref{NDFcooling} for each of the four paths. This function represents the average number of Eu$^3+$ ions in a coordination shell as a function of the distance from the charged O$^-$ specie of the MP functional group. Each plot in these figures includes a family of colored neighbor distribution function curves corresponding to different times along the paths of Figure~\ref{Thermodynamic_Paths}. The neighbor distribution function curves for the initial states are colored blue, and the distribution function curves for the final stable equilibrium states are the red curves.

For Paths I and II (Figure~\ref{NDFheating}), the number of Eu$^{+3}$ ions a distance of 4 \AA $\:$ from a MP functional group initially has a maximum $n_d(r)$ of over 2.5 (noting that the green curve overlays the blue curve of the initial state so that it is not seen in Figure \ref{NDFheating}(a)) and 2.2, respectively, and decreases to less than 1.0 when mutual stable equilibrium with the high-temperature reservoir is reached. In other words, Eu$^{+3}$ is removed from MP functional sites along the polymer chain as energy is added to the system during heating.  For the case of Path II, the Eu$^{+3}$ ions are almost completely driven away from the functional group when the system reaches the point of maximum energy (i.e., when the maximum $n_d(r)$ is close to zero). With cooling, the final maximum $n_d(r)$ returns to 1.0 as the energy drops and the sequestration subsystem comes to final mutual stable equilibrium with the high-temperature reservoir.

Conversely, for the two cooling paths (Paths III and IV), the maximum $n_d(r)$ for the number of Eu$^{+3}$ ions a distance of 4 \AA $\:$ from an MP functional group increases, as seen in Figure~\ref{NDFcooling}, from less than 0.4 for both paths to 2.5 when mutual stable equilibrium is established at the low-temperature reservoir. This behavior corresponds to the sequestration of Eu$^{+3}$ ions from the aqueous solution by the MP functional groups. Interestingly, although the final $n_d(r)$ for the Eu$^{+3}$ is the same for these two cooling paths, the times over which Eu$^{+3}$ ions are captured are quite different. Sequestration takes place rapidly along Path III, but along Path IV, the Eu$^{+3}$ quickly reaches a  coordination value of around 1.6, pauses, and then slowly reaches the maximum value of 2.5.  This two-stage capture process demonstrates how the thermodynamic path taken during cooling affects the {\em rate} of sequestration. 

\begin{figure}[ht!]
	\begin{center}
		\includegraphics[width=0.45\textwidth]{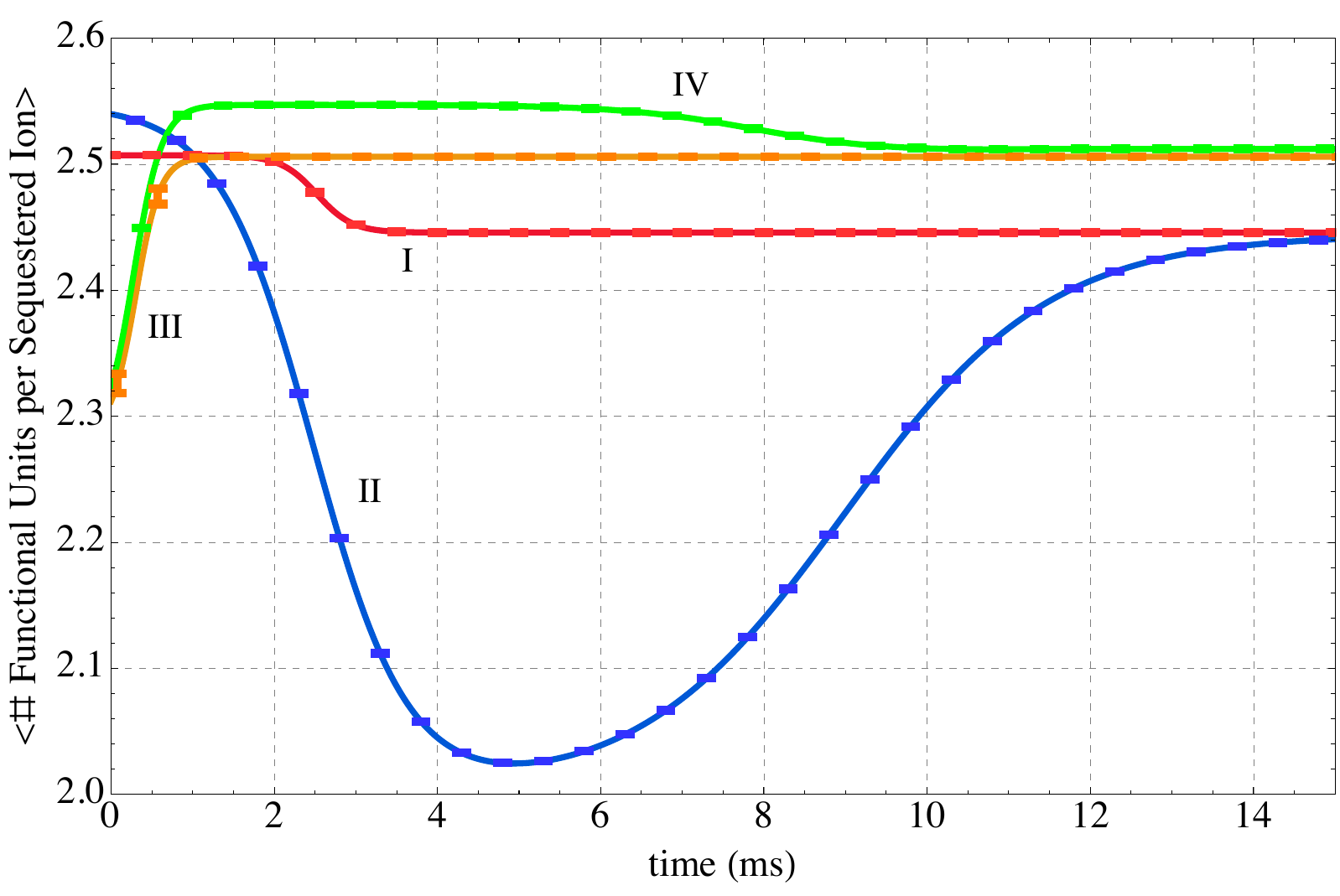}
  \hspace*{\fill} 
		\caption{Expected number of interacting functional units per sequestered ion along the four kinetic paths defined in Figure~\ref{Thermodynamic_Paths}. The sequestration subsystem is heated along Path I, heated and cooled along Path II, and cooled along Paths III and IV. }
		\label{Num_Seq}
	\end{center}
\end{figure}

\begin{figure}[ht!]
	\begin{center}
		\includegraphics[width=0.45\textwidth]{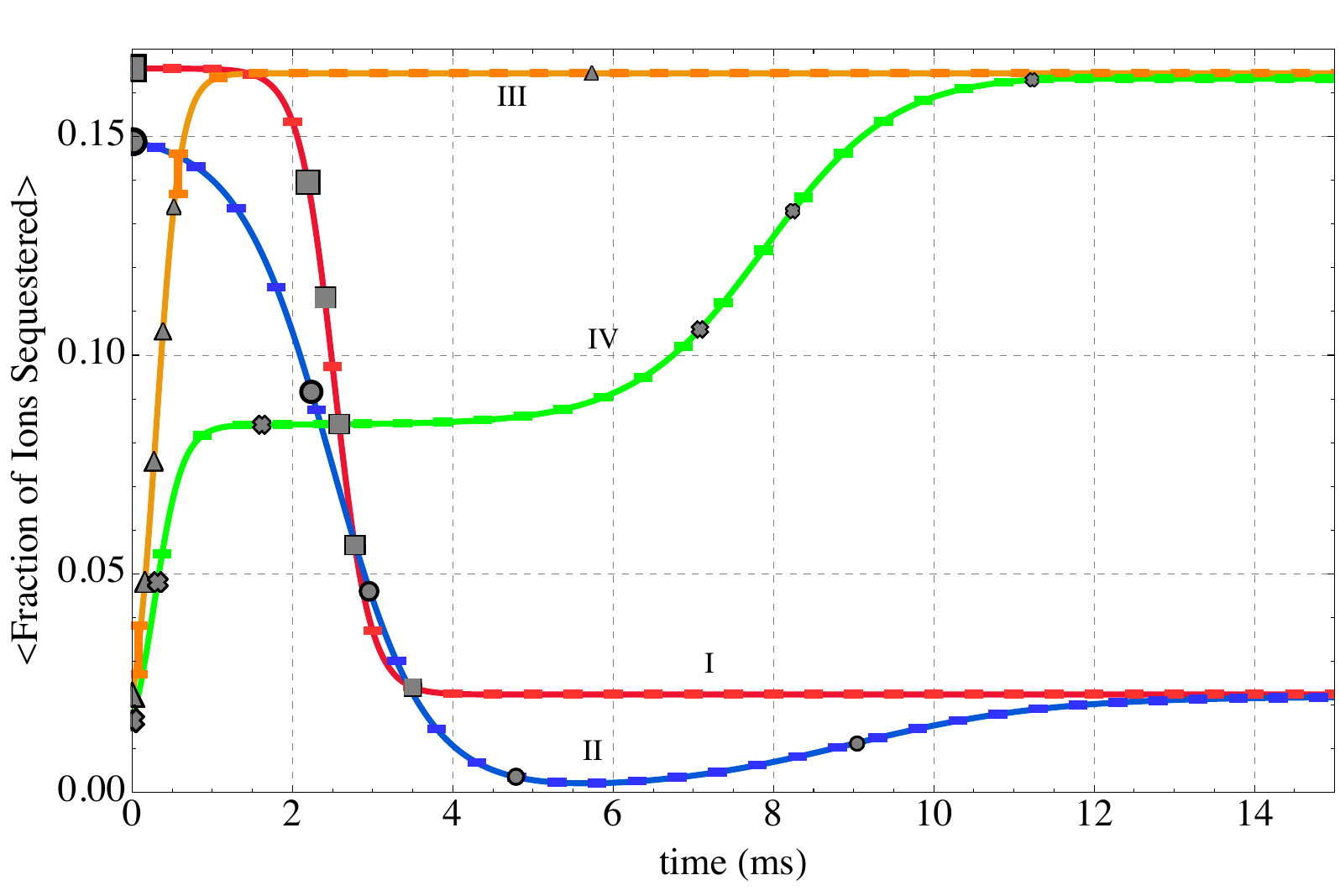}
  \hspace*{\fill} 
		\caption{The expected number of sequestered ions in the system along the four kinetic paths defined in Figure~\ref{Thermodynamic_Paths}. The sequestration system is heated along Path I, heated and cooled along Path II, and cooled along Paths III and IV. }
		\label{SeqEff}
	\end{center}
\end{figure}

The extent to which the MP functional groups are utilized to capture Eu$^{+3}$ ions is given by the overall expected fractional units per sequestered Eu$^{+3}$ shown in Figure~\ref{Num_Seq}. The overall expected number of sequestered ions appears in Figure \ref{SeqEff}.  Both of these expectation quantities are based on arithmetic averages for individual energy levels and the probability distributions predicted by the SEAQT equation of motion along each of the four thermodynamic paths. As seen in Figure~\ref{Num_Seq}, the utilization of the MP functional groups in capturing Eu$^{+3}$ ions is quite similar for the four non-equilibrium thermodynamic paths.  As to the number of sequestered Eu$^{+3}$ ions, which is an indication of the sequestration performance, two observations can be drawn from Figure~\ref{SeqEff}. The first is that the fraction of sequestered Eu$^{+3}$ ions  follows a trend inverse to the system energy.  For example, Paths I and II begin at low energies and increase to higher final energies and the fraction of sequestered ions progresses from high value near 0.15 to a low value near 0.02.  Similarly,  energy decreases along the cooling paths while Eu$^{+3}$ sequestration increases. The second observation is that just as the overall fraction of sequestered Eu$^{+3}$ ions at stable equilibrium is related to the subsystem energy so are the instantaneous fraction of sequestered ions along the paths connected to the instantaneous subsystem energy.  This observation can be seen by comparing Paths III and IV in Figure~\ref{SeqEff}. Both curves reach the same final fraction of sequestered ions (because they have almost identical final energies), but sequestration along Path III occurs relatively quickly whereas the Eu$^{+3}$ ions are captured in two stages along Path IV.  The two-stage nature of sequestration along Path IV follows the same two-stage evolution of energy  along this path that is evident in Figure~\ref{EP_1}. 

\section{Discussion}\label{Discussion}

The four thermodynamic paths considered in this study suggest the amount of sequestered Eu$^{+3}$ is closely related to the system energy. However, the time-evolution of the system energy is determined by the steepest-entropy-ascent principle via the SEAQT equation of motion. For this reason, entropy production ultimately determines the rate of sequestration.

Since the kinetic path obtained by solving the SEAQT equation of motion depends upon the initial state of the system and the reservoir temperature, it is possible to choose different conditions that control the overall sequestration behavior. Two categories of behavior were observed. Paths I and III produced a monotonic release or absorption of Eu$^{+3}$ ions during heating or cooling, respectively, while Paths II and IV exhibited more complicated processes of sequestration. Paths I and III have sigmoidal sequestration kinetics (either Eu$^{+3}$ ion release during heating or absorption during cooling). 

Paths II and IV, on the other hand, exhibit stages or transitions in sequestration kinetics. For example, during heating along Path II, initially sequestered Eu$^{+3}$ ions are released and steepest entropy ascent steers the system towards a maximum energy. At this point, all the Eu$^{+3}$ have been rejected from PEI-MP into the solvent. This process is manifested by the maxima or minima in the blue curves in Figures~\ref{EP_1}, \ref{Rg_t_1}, \ref{Num_Seq}, and \ref{SeqEff}. The absence of sequestered Eu$^{+3}$ ions at this stage is evident in conformation II$_4$ of Figure~\ref{RepMicro_1}. With the beginning of cooling along Path II, some Eu$^{+3}$ ions in the solvent are re-captured by the PEI-MP network.  Cooling along Path IV also exhibits multiple stages.  In this case, a substantial number of Eu$^{+3}$ ions are quickly sequestered during cooling, then absorption pauses for a period of time before resuming and continuing to the stable equilibrium sequestered concentration. This kind of pause along a kinetic path is reminiscent of a kinetic pause first noted by Berreta~\cite{Beretta2006} in a modeled system as it passed near a metastable equilibrium state on its way to stable equilibrium. At this point, it is worth reiterating that the thermodynamic paths shown in Figure~\ref{Thermodynamic_Paths} are not drawn arbitrarily but rather are determined by the equation of motion using the steepest-entropy-ascent principle. Thus, it seems plausible that Paths II and IV follow a nonlinear trajectory in Figure~\ref{Thermodynamic_Paths} because maximum entropy production directs the sequestration system toward some unapparent metastable state that acts like an attractor along the path to stable equilibrium.

As a final note, scaling of the dimensionless time produced by the SEAQT equation utilizes the calculated change in the radius of gyration, the expected literature-derived polymer diffusion constant, and a proportionality constant taken from a previous study of polymer folding systems with a similar combined approach of REWL and SEAQT~\cite{McDonald2023polymer}. The dimensionless time evolution for Path I takes approximately 5 time steps, the calculated change in $R_g$ is 10 \AA, and an approximate diffusion constant for the polymer of 10$^{-6}$ cm$^2$/s is used. Utilizing the equations and proportionality constant from~\cite{McDonald2023polymer} provides a scaled evolution time of approximately 4.5 ms.

\section{Conclusions}\label{Conclusions}   

The contribution presented here uses the SEAQT framework to predict the sequestration kinetics of Eu$^{+3}$ ions in water to polyethyleneimine functionalized with methylenephosphonic acid. The SEAQT equation of motion is applied to an energy eigenstructure constructed with the Replica Exchange Wang-Landau algorithm using the Bond Fluctuation Model and standard pair-potentials. The following conclusions are drawn from the predicted sequestration behavior along four different non-equilibrium heating and cooling paths.

\begin{enumerate}
\item The fraction of Eu$^{+3}$ ions sequestered  is sensitive to the total energy of the sequestration subsystem. The lower the energy, the higher the fraction of sequestered Eu$^{+3}$ ions is.

\item The number of sequestered Eu$^{+3}$ ions is sensitive to the system energy.  Since the reservoir temperature establishes the final energy, temperature is the most important factor controlling the number of sequestered Eu$^{+3}$ ions.

\item The steepest-entropy-ascent principle determines whether the thermodynamic path follows a direct route to stable equilibrium or a more convoluted path with multiple stages (like Paths II and IV). We speculate that paths with multiple stages arise from metastable equilibrium states that lie near the kinetic path taken.

\item For the four non-equilibrium thermodynamic paths considered here, Path III captures the most Eu$^{+3}$ ions in the shortest time. This path starts at an initial high-energy non-equilibrium state not far removed from the locus of stable equilibrium states with energy extracted in a heat interaction to a low-temperature reservoir.  

\end{enumerate}

\section{Acknowledgements}
JM acknowledges support from the U.S.~Department of Education through the Graduate Assistance in Areas of National Need Program
(grant number P200A180016).

\bibliography{JaredsRefs_Seq.bib} 

\end{document}